\PassOptionsToPackage{table,xcdraw}{xcolor}
\documentclass[acmsmall,prologue,table,authorversion]{acmart}

\AtBeginDocument{%
  \providecommand\BibTeX{{%
    \normalfont B\kern-0.5em{\scshape i\kern-0.25em b}\kern-0.8em\TeX}}}

\setcopyright{rightsretained}
\acmJournal{TOCHI}
\acmYear{2024} \acmVolume{1} \acmNumber{1} \acmArticle{1} \acmMonth{1}\acmDOI{10.1145/3689044}

\usepackage[utf8]{inputenc}
\usepackage[T1]{fontenc}
\usepackage{import}
\usepackage{booktabs}
\usepackage{textalpha} 
\usepackage[super]{nth}
\usepackage{float}
\usepackage{lscape}
\usepackage{graphicx}

\usepackage{comment}
\usepackage{xspace}
\usepackage{color}

\usepackage{xargs}                      

\definecolor{applegreen}{rgb}{0.55, 0.71, 0.0}
\definecolor{amber}{rgb}{1.0, 0.49, 0.0}
\definecolor{fuschia}{rgb}{0.57, 0.36, 0.51}
\definecolor{ashgrey}{rgb}{0.7, 0.75, 0.71}
\definecolor{beaver}{rgb}{0.62, 0.51, 0.44}
\definecolor{teal)}{rgb}{0.0, 0.5, 0.69}
\definecolor{blue-violet}{rgb}{0.54, 0.17, 0.89}

\begin{document}

\title{Beyond Intrinsic Motivation: The Role of Autonomous Motivation in User Experience}


\author{Dan Bennett}
\email{pavementsands@gmail.com}
\orcid{0000-0002-9330-5529}
\affiliation{%
  \institution{University of Bristol}
\country{UK}}
\author{Elisa D. Mekler}
\orcid{0000-0003-0076-6703}
\affiliation{%
  \institution{IT University of Copenhagen}
\country{Denmark}
}


\begin{abstract}
Motivation and autonomy are fundamental concepts in Human-Computer Interaction (HCI), yet in User Experience (UX) research they have remained surprisingly peripheral.
We draw on Self-Determination Theory (SDT) to analyse autonomous and non-autonomous patterns of motivation in 497 interaction experiences.
Using latent profile analysis, we identify 5 distinct patterns of motivation in technology use --- ``motivational profiles'' --- associated with significant differences in need satisfaction, affect, and usability. Users' descriptions of these experiences also reveal qualitative differences between profiles: from intentional, purposive engagement, to compulsive use which users themselves consider unhealthy. 
Our results complicate exclusively positive notions of intrinsic motivation, and clarify how extrinsic motivation can contribute to positive UX.
Based on these findings we identify open questions for UX and SDT: addressing ``hedonic amotivation'' --- negative experiences in activities which are intrinsically motivated but not otherwise valued --- and ``design for internalisation'' --- scaffolding healthy and sustainable patterns of engagement over time.

\end{abstract}





\maketitle

\section{Introduction}



Our \textit{motivation} for an activity concerns the psychological processes and factors that mould our engagement with that activity \cite[][p.~45]{sheldon_self-determination_2008}. This includes not only our goals, but also our relation to the activity, how this develops and is influenced by external factors \cite[][p.~181]{ryan_self-determination_2017}.
Motivation influences a range of outcomes which HCI research has considered important, including emotion \cite{liu_motivational_2022}, depth of engagement \cite{fryer_understanding_2016}, effort, and achievement \cite{wang_latent_2017}. Perhaps naturally, then, motivation has been considered a central issue in HCI research \cite{ballou2022self}: it plays an important theoretical role 
in UX and design work \cite[e.g.][]{gerstenberg_designing_2023, partala_understanding_2012,peters_designing_2018}, and is implicated in many of the field's ``grand challenges'' \cite{stephanidis_seven_2019, shneiderman_grand_2016}. 
However, despite this acknowledged importance
there remain gaps in our understanding of motivation in technology use. It is unclear how motivation influences user experience (UX), and how technologies and contexts in turn influence motivation.

Insofar as UX research has addressed the issue of motivation, it has commonly focused on the motivating role of immediate \textit{satisfactions} that arise during interaction
\cite[e.g.,][]{hassenzahl_needs_2010, hassenzahl_experience-oriented_2015, mekler_momentary_2016, tuch_leisure_2016}. This approach, grounded in Self-Determination Theory (SDT) \cite[see e.g.,][]{sheldon_what_1996}, addresses how interactions satisfy psychological needs, and how this in turn influences positive affect, product evaluation, and other valued outcomes.  In particular, SDT posits that three ``basic psychological needs'' ---  autonomy, competence and relatedness --- are expected to promote \textit{intrinsic} motivation: spontaneous enjoyment, interest, and continued engagement with the activity \cite[][p.~123-178]{ryan_self-determination_2017}.
SDT considers intrinsic motivation inherently autonomous, due to its basis in spontaneous interest \cite[p.~14] {ryan_self-determination_2017}, and also associates it with learning and growth \cite[p.~222]{shaver_beyond_2012}. In line with this, HCI and UX research has often treated intrinsic motivation as the gold standard of motivation \cite[see e.g.,][]{alexiou_digital_2018, song_effects_2013,lushnikova_self-determination_2023, tyack_self_2024}.
However, while SDT 
grants an important role to intrinsic motivation, it holds that behaviour in adult life is at least as strongly influenced by \textit{extrinsic} motivations, motivations which arise not from the activity itself but from its outcomes and wider ramifications \cite[][pp.~196-7]{ryan_self-determination_2017}. Like intrinsic motivation, extrinsic motivation \emph{can} in many cases be highly autonomous
\cite[][p.~198]{ryan_self-determination_2017}. It can also be strongly supportive of engagement
\cite{fryer_understanding_2016}, achievement \cite{wang_latent_2017} and other positive outcomes \cite[][p.~198]{liu_motivational_2022, fryer_understanding_2016, gustafsson_motivational_2018, ryan_self-determination_2017}. 
%
%
Despite all this, extrinsic motivation has commonly been neglected in HCI and UX research \cite[e.g.][]{song_effects_2013, alexiou_digital_2018}, or else reduced to its less healthy and autonomous forms (such as reward and punishment) and exclusively associated with poor quality outcomes \cite{turkay_self-determination_2023, hammer_quality--user-experience_2018}. 
This leaves gaps in our understanding of users' organic motivations, how users come to value their interactions, and factors which support meaningful UX.

To address this, we present a study grounded in
Organismic Integration Theory (OIT, \citep{ryan2017organismic}) --- a sub-theory of SDT. 
OIT identifies six distinct forms of 
motivation, associated with different relationships to the self, qualities of experience, degrees of autonomy, and patterns of behaviour \citep{ryan2017organismic}.  
We conduct an online survey (n=497) addressing participants' experiences with technologies they use regularly, examining the role of motivation in user experience. 
Building on recent developments in motivational psychology \citep{howard2016motivation}, we use statistical modelling techniques to 
identify distinct \textit{motivational profiles} in our sample which capture the relative influences of OIT's 6 motivational regulations.
We then connect these profiles to distinct patterns of user experience, captured via UX measures, as well as participants' own descriptions of their experience. 



The contribution of our work is fourfold. First, our findings push beyond the focus on need satisfaction prevalent in UX research \cite[e.g.,][]{hassenzahl_needs_2010,peters_designing_2018,tuch_analyzing_2013}, to clarify the ways in which motivational regulations impact user experience. We identify five distinct ``motivational profiles'' in technology use and find that they are associated with significant differences in basic need satisfaction, affect, and perceived usability. 
Second, our work extends understandings of extrinsic motivation in UX: 
Our results clarify the conditions under which extrinsic forms of motivation are autonomous and contribute positively to UX, 
dispelling the idea that extrinsic motivation in technology use can be reduced to a singular, negative phenomenon. 
Third, we bring nuance to the view, common in SDT and UX research, that intrinsic motivation is unequivocally positive \cite{ballou2022self}. We identify what we call \textit{hedonic amotivation}: intrinsically motivated, yet low-intentionality, experiences which are marked by compulsion, and which users themselves consider unhealthy. We discuss the conceptual challenge this state poses for SDT theory, and identify future work to understand it.
Finally, consolidating our findings with prior work on  
 motivational design \cite{gerstenberg_designing_2023}, 
we articulate a research agenda for \textit{design for internalisation} to support users in developing autonomous and sustainable motivation.

\section{Background}

\begin{figure*}
\centering
\includegraphics[width=\columnwidth]{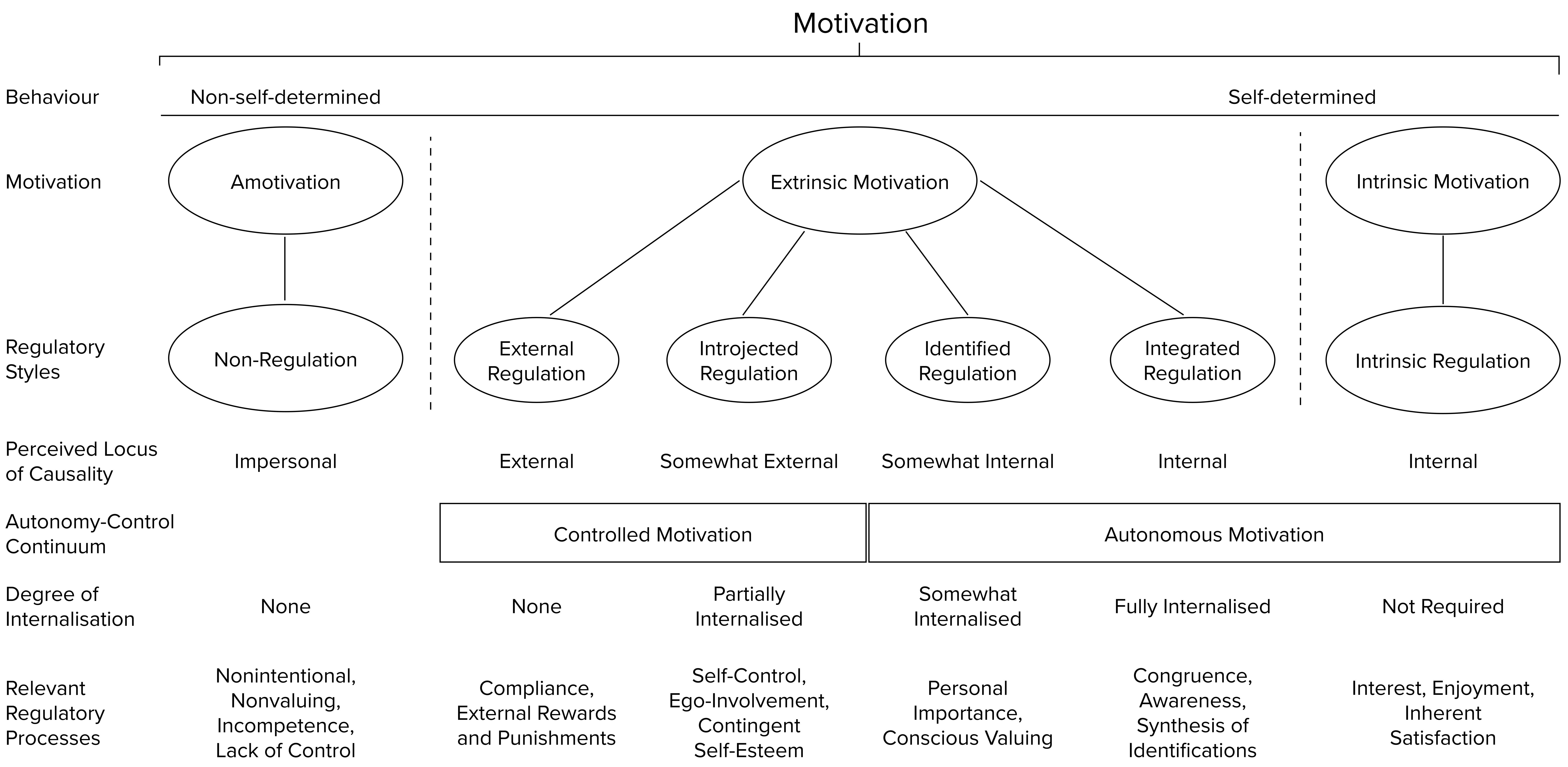}
\vspace{-1em}
\caption{Motivation as conceptualised by Self-Determination Theory. 
Organismic Integration Theory (OIT) posits 6 ways in which motivation can be regulated, ranging from the least self-determined and autonomous (amotivation) to the most self-determined and autonomous (integrated regulation and intrinsic motivation). Note: the placement of intrinsic motivation to the right of integrated regulation does not indicate it is \emph{more} autonomous. Adapted from \protect\citep{ryan2017organismic, vansteenkiste_fostering_2018}.}~\label{OIT_diagram}
\vspace{-2em}
\end{figure*}

\subsection{Characteristics of Good and Bad User Experience}
\label{bg_goodbad_ux}
In the last decade much quantitative UX research has followed a template set by Hassenzahl and colleagues
\cite{hassenzahl_needs_2010}: focusing on how technologies satisfy users' psychological needs, and how this in turn relates to product perception, affect and other variables of interest.
Much of this work \cite[e.g.][]{hassenzahl_needs_2010, hassenzahl_experience-oriented_2015, mekler_momentary_2016, tuch_leisure_2016} has focused on the satisfaction of ten ``candidate psychological needs'' identified and evaluated by Sheldon and colleagues \cite{sheldon_what_2001}. These range from the need for security, to the need for money and luxury. 

This approach to UX in terms of need-satisfaction is grounded in Self-Determination Theory (SDT), where three needs in particular are granted special status: autonomy (the need to self-regulate one’s experiences and actions), competence (the need to feel effectance and mastery), and relatedness (the need to feel socially connected, belong, and be significant to others). 
SDT theorises that these three ``basic'' needs are central to human behaviour and thriving, and supportive of positive affect \cite[p.~242]{ryan_self-determination_2017}; an idea supported by a large and growing body of work \cite{ryan_we_2022}, and by the eminence of these three needs in Sheldon et al.'s results \cite{sheldon_what_2001}.
In UX research, however, these three basic needs have not always received a level of attention commensurate with their status in SDT. In particular, much work has excluded autonomy from analysis in favour of other less theoretically central candidate needs \cite[e.g.,][]{hassenzahl_needs_2010, hassenzahl_experience-oriented_2015, mekler_momentary_2016, tuch_leisure_2016}.
This exclusion is striking. Within SDT, autonomy has a ``special status'' as theoretically ``central'' even among the three basic needs \cite[p.~97]{ryan_self-determination_2017}, while in wider HCI autonomy has long had a central role \cite{bennett_how_2023}, being considered key in many of the field's grand challenges \cite{stephanidis_seven_2019, shneiderman_grand_2016}. 

Another concept which is central in SDT but often neglected in UX research is motivation.
While there are multiple accounts of motivation in psychology \cite[e.g.,][]{sheldon_wanting_2011, rheinberg_intrinsic_2018, hattie_theories_2020}, one inclusive definition is that it concerns ``the causes or reasons that underlie a given behavior'' \cite[][p. 286]{kleinginna_categorized_1981}.
This does not only concern goals --- what we want to achieve --- but the way 
intentional and purposeful behaviour is moulded by psychological processes and contextual factors \cite[][p.~45]{sheldon_self-determination_2008}.
In SDT these psychological and contextual factors are divided into \textit{intrinsic motivation} (following from the activity itself) and \textit{extrinsic motivation} (motivation related to factors outside of, or consequent upon the activity itself). 
Both forms of motivation can make positive contributions to well-being and experience \cite[pp.~179-183]{ryan_self-determination_2017}, and both are influenced by basic need satisfaction. Intrinsic motivation is expected to arise \textit{immediately} from basic need satisfaction during the activity itself. Meanwhile basic needs play 
a less direct, but still crucial role in extrinsic motivation: shaping the character of motivation over longer time periods, by
providing the conditions which help us connect activities to our identities, values and goals
\cite[pp.~184-190]{ryan_self-determination_2017} (see \autoref{sec:bg_OIT}).  

When HCI and UX research has drawn on SDT it has tended to focus almost exclusively on need satisfaction and its impact on intrinsic motivation \cite[e.g.][]{alexiou_digital_2018, burgers_how_2015, kao_effects_2018, turkay_self-determination_2023}. Extrinsic motivation, meanwhile, has either been neglected \cite[e.g.][]{song_effects_2013, alexiou_digital_2018}, or else treated as a straightforwardly lesser and even negative form of motivation 
\cite{turkay_self-determination_2023, hammer_quality--user-experience_2018}.
Exceptions to this are relatively rare. A recent paper on the Technology Acceptance Model,
addresses how intrinsic and extrinsic motivation support the acceptance of technologies \cite[e.g.][]{stiegemeier_motivated_2022}, though the relationship of acceptance to UX remains under-specified \cite{hornbaek_technology_2017}.
Elsewhere, Mekler and Hornbæk's work on eudaimonic and hedonic UX \cite{mekler_momentary_2016} makes a broad distinction between goal-type in motivation between immediate pleasure from more meaning-oriented goals. 
However, such work does not address the user's relationship to the goal - why it matters, and whether the motivation following from the goal is autonomous --- important issues in SDT's conception of motivation.

Recent work has articulated an approach to motivation and basic need satisfaction in UX which is more explicitly grounded in SDT. The Motivation, Engagement and Thriving in User Experience (METUX) framework \cite{peters_designing_2018} moves beyond previous SDT-grounded UX research in emphasising the way need satisfaction can vary in different spheres of engagement: for example, at the interface, in wider behaviour, or in life. Work developing out of this framework has provided evidence of differing patterns of need satisfaction across these different spheres, pointing to a complexity in the effect of need satisfaction that has not been addressed in previous UX work \cite{burnell_technology_2023}. 
While the METUX account does describe the positive contribution of extrinsic motivation to UX and well-being outcomes, to date this account has remained theoretical, and articulated at a relatively high level. Meanwhile empirical work grounded in the framework has not addressed extrinsic motivation \cite{burnell_technology_2023}.




\subsection{Organismic Integration Theory} \label{sec:bg_OIT}

Both the \emph{quality} and \emph{quantity} of motivation influence its capacity to energise and direct behaviour \citep{ryan2000intrinsic}. 
\emph{Organismic Integration Theory} (OIT) is a sub-theory of SDT which proposes a model of motivational \textit{quality} based on six different 
ways in which motivation can be regulated. These six  \textit{motivational regulations} are qualitatively distinct, 
yet can be ordered in terms of their relative level of autonomy (see \autoref{OIT_diagram}).

Four of these regulations concern extrinsic forms of motivation: behaviours performed for reasons other than the inherent interest or enjoyment of the activity. Extrinsically motivated behaviours are autonomous insofar as they are coherently organised with other facets of the self, through \emph{internalisation}. Internalisation is a process whereby activities and behaviours become connected to goals, values and other aspects of the self. OIT holds that this process is facilitated by contexts which satisfy the three basic needs discussed above \citep{deci1994facilitating}. 
The most autonomous form of extrinsic motivation is 
\emph{integrated regulation}, wherein behaviours are fully \textit{integrated} with the self, becoming congruent with other activities and goals, and valued in a reflective and intentional manner.
\textit{Identified regulation} is also autonomous, but does not involve such strong congruence with identity. Rather, the activity and its outcomes are identified as being personally important, yet can remain compartmentalised with respect to other aspects of the self. I may, for example, consider it valuable to run regularly to maintain my health and energy levels, without forming a strong personal connection to the activity, or identifying as a sporty person.

Contrasting with these two autonomous forms of regulation are two more 
\textit{controlled} forms of regulation.
First, \textit{introjected regulation} concerns values and behaviours which are subject to rigid internal controls such as guilt, shame, or contingent self-esteem \citep{deci1994facilitating}. Although introjected regulations are partly internalised, they remain controlling since their rigidity puts them in conflict with other values and beliefs. 
The most controlled form of extrinsic regulation is \emph{external regulation}. This concerns behaviours which are not internalised but instead ``motivated by and dependent upon external reward or punishment contingencies'' \citep[p.~184]{ryan2017organismic}. 
Finally, these four forms of extrinsic motivation are joined by \textit{intrinsic motivation} --- motivation following from the \textit{intrinsically} ``engaging and fascinating'' nature of the activity \cite[][p.~222]{shaver_beyond_2012}, and \emph{amotivation} --- behaviour performed without intentionality. 

While some work in HCI has treated intrinsic motivation as the most autonomous and well-being supportive form of motivation \cite[e.g.,][]{hammer_quality--user-experience_2018, song_effects_2013, turkay_self-determination_2023}, this does not find a basis in SDT which emphasises that integrated extrinsic regulation is no less autonomous \cite[e.g.,][p.~197]{ryan_self-determination_2017}. 
Studies that examine 
integrated regulations alongside intrinsic motivation 
\citep{burton_differential_2006,mankad_motivational_2014,murray_mechanisms_2020,sheehan_associations_2018} indicate that both these forms of autonomous motivation predict beneficial psychosocial outcomes \citep[e.g., lower anxiety;][]{sheehan_associations_2018} and behaviours \citep[e.g., engagement;][]{mankad_motivational_2014}.
The difference between these two forms of motivation is, instead, functional. Unlike (extrinsically motivated) integrated activities, intrinsically motivated activities \textit{do not} rely on a connection or association with values, goals and identity in our wider life.
Intrinsically motivating activities instead motivate via satisfactions \textit{inherent to the activity itself}: they are ``engaging and fascinating'' \cite[p.~222]{shaver_beyond_2012} for reasons which need not relate to the user's reflective goals and values. 
Connected to this functional difference, experiences of extrinsic and intrinsic motivation have different temporal properties. Since extrinsic motivations relate to consequences and outcomes of an activity, they may not be experienced strongly during the activity itself. Intrinsic motivations, meanwhile, are typically spontaneous, experienced during the activity itself, 
 and related to the person's present concerns and impulses \cite[p.~198]{ryan_self-determination_2017}. 

There is evidence that the different motivations and regulations posited by OIT are not mutually exclusive \citep{howard2016motivation} but typically operate together. A person can be motivated to put effort into an activity for multiple reasons simultaneously. They may seek financial gain and the approval of others, but also feel the activity is aligned with personal values, and immediately enjoyable. OIT holds that the overall autonomy of this motivational mix is a key factor in wellbeing and outcomes. Initially, SDT research measured the autonomy of motivation by constructing a \emph{Relative Autonomy Index} \citep{grolnick1989parent}, which ``approximates an individual's position along the underlying continuum of self-determination'' \citep[p.~536]{howard2020review} using weighted self-report data. 
However, this approach has been criticized since 
``the weights associated with each [motivation] subscale are relatively arbitrary with no published empirical evidence to support them'' \citep[p.~536]{howard2020review}.
More fundamentally, studies have shown that each of the different regulations contribute in a qualitatively differently way to behaviour, experience and outcomes \cite[e.g.,][]{koestner_identified_1996,weinstein_when_2010,shaver_beyond_2012}. When we reduce motivation to a one-dimensional score for relative autonomy, we lose sight of all this. As such, more granular approaches have been developed, employing 
statistical modelling methods such as latent profile analysis which allow attention to the contributions of all six motivational regulations \citep[e.g.][]{fernet2020self,howard2018using,howard2021longitudinal}. 

Until recently there has been relatively little engagement with OIT in HCI research. 
One exception to this is found in work by Gerstenberg et al., which draws on OIT theory to guide the design of behaviour change technologies. The authors argue that in the context of behavior change intrinsic motivation is less relevant than autonomous extrinsic motivation \cite{gerstenberg_designing_2023, gerstenberg_designing_2024}. 
Elsewhere, an OIT-based psychometric scale was developed as part of the METUX framework \citep{peters_designing_2018}, though this addresses a limited set of regulations --- excluding amotivation and integrated regulation. 
Another scale, the User Motivation Index (UMI) \cite{bruhlmann_measuring_2018}, addresses all six regulations in the context of technology use. This has recently been used with latent profile analysis to study motivational profiles in multiplayer online gaming, clarifying how intrinsic and extrinsic motivations come together to shape patterns of need satisfaction and player experience \cite{bruhlmann_motivational_2020}.
Other work using the UMI has analysed the motivational regulations in a more piecemeal fashion. For example: work on museum websites found a positive correlation between intrinsic motivation and needs satisfaction, and observed differences in introjection between different museum websites \cite{lushnikova_self-determination_2023}, while 
work on in-vehicle technologies found that amotivation and identified regulations 
explained variance in autonomy satisfaction \cite{stiegemeier_user_2022} and the intention to use them \cite{stiegemeier_motivated_2022}.


\section{Methods}

To understand how user experience 
is influenced by intrinsic and extrinsic motivations
we conducted an online survey, then employed 
latent profile analysis (LPA), to identify multi-dimensional profiles of user motivation across Organismic Integration Theory's (OIT) six motivational constructs. 

Each study participant was asked to describe their experience with a technology they either ``felt good'' about (positive condition) or ``did not feel good about (negative). This two-condition design was not intended to serve two purposes. 
First, to ensure our results reflected motivation across a representative range of both positive and negative experiences, and second to allow comparison 
with previous studies which have investigated the relationship between need satisfaction and positive UX \cite[e.g.,][]{hassenzahl_needs_2010,tuch_analyzing_2013}. We did not, however, form or test hypotheses about differences between these two conditions.

Our sample size was guided by prior literature and a pilot analysis. Power analysis for LPA faces a priori uncertainty about population parameters and class-counts. 
As such it is recommended to select sample size based on empirical evidence \cite{spurk_latent_2020, dalmaijer_tutorial_2023}. 
We conducted a pilot analysis of a prior UMI dataset containing 460 participants \cite{bruhlmann_measuring_2018}, observing that the data best supported a 4-class structure with unequal class sizes. This solution resulted in strong model fits, and when comparing outcomes across profiles, we observed medium-to-large effect sizes.
We supported our sample size selection by consulting Monte-Carlo simulation studies. Nylund et al. found that for a similar unequal 4-class solution a sample size of 200 resulted in identification of the correct number of classes 100\% of the time \cite{nylund_deciding_2007}. As such we aimed for a sample size of approx. 500 participants\footnote{Subsequent to our analysis more precise guidelines have been published \cite{dalmaijer_tutorial_2023}, which confirm the adequacy of this sample size for our observed group sizes and feature count.}.


Our study combines confirmatory testing with theory-informed exploratory analysis, consistent with recent recommendations concerning good research practice \cite{scheel_why_2021, van_lissa_developmental_2022}.  The confirmatory aspect of our study tests the following hypothesis:
\begin{itemize}
    \item[] H1: Distinct motivational profiles (patterns across OIT's 6 motivational regulations) will be associated with significant differences across the three basic needs, perceived usability, affect (PANAS), and self-attribution. 
\end{itemize}

Alongside the hypothesis variables, we capture additional data to support exploratory analyses --- needs beyond the three basic needs and free-text responses. These support further interpretation of the motivational profiles and of their relationship to the hypothesis variables, and help identify directions for future work.

In the remainder of this section, we describe our sample and study design.  Data, analysis code, and preregistration (as time-stamped wiki) are provided at 
\url{https://osf.io/eny6v/}.

\subsection{Participants}
Participants were recruited 
via \url{prolific.co}, a crowdsourcing platform. Prolific's pre-screen options were used to limit participants to those over 18, fluent in English, and holding a 100\% approval rating over a minimum of 3 previous studies. In total, we collected 573 full responses. 

These responses were then screened for data quality. 
For example, we checked responses to the open questions to ensure participants followed instructions: e.g., that they focused on direct personal experience rather than broad opinion. 
We also investigated statistical markers of careless responding using the ``careless'' package in R, calculating ``synonym'' and ``antonym'' checks, even-odd consistency, and the Mahalonobis distance \cite[see][]{yentes_careless_2018}. Participants flagged on one or more checks were manually investigated for wider patterns of carelessness (e.g., multiple choice answers strongly inconsistent with the open question response, or inconsistent responding across closely related items). 
Participants were also excluded if they responded incorrectly to at least one of three attention check questions. 
The exclusion decision process is fully documented in the code available in the \href{https://osf.io/eny6v/}{OSF repository}. 
After exclusions (n = 76), N = 497 participants remained for analysis. 
The median age was 28 ($min = 18; max = 73$), 250 indicated they were male, 235 female, 6 non binary, and 1 transgender male. Three gave unclear answers (e.g. ``hetero''), and two rejected the question.




\subsection{Procedure}
Upon clicking the survey link in Prolific, participants were introduced to the study and data handling procedure. After giving consent, participants were asked to provide demographic information, 
before proceeding to 
the study. 

Participants were randomly assigned to one of two conditions, describing a digital technology they either ``felt good'' about (positive condition) ($n=254$) or ``did not feel good'' about (negative condition) ($n=243$). Participants were told to understand "feel good" in whatever way made sense to them, and to focus on a technology they used frequently over the past two months.
They were guided to think of a single, specific, technology such as a particular app, website, software, gadget, etc., and not a general category such as "computers" or "phones". To confirm that participants understood these criteria, they
answered a multiple choice question demonstrating understanding of the criteria, and were directed back to the instructions until they answered correctly.
They were then asked to name the technology, report how regularly they used it, and spend a few moments thinking about their experiences with this technology before continuing. 

Next, participants were asked to describe their experience with the technology in their own words. To support descriptive writing,
we followed recommendations to structure the writing with multiple prompts \cite{tuch_analyzing_2013}, basing these on prompts used successfully in previous studies \cite{mekler_momentary_2016, tuch_analyzing_2013, bruhlmann_measuring_2018}. 
First, participants were asked why they [do / do not] feel good about the technology. They were instructed to describe this as accurately, and in as much concrete detail as they could, writing at least 50 words. They were told to focus on their own personal experience with the technology, and how it fits into their life, thinking about their overall experience with the technology, not just one occasion. 

This was followed by three more specific questions to draw out particular details relevant to motivation. (1) \textit{Why did you \textit{start} using [\textit{technology name}]}; (2) \textit{Why do you \textit{continue} to use [\textit{technology name}]}; (3) \textit{What does [\textit{technology name}] mean to you? How does it relate to your values and goals?}. Through all these questions, participants were told to write in their own words without external aids (including LLMs). Copying and pasting into free text fields was disabled. 
At the end of the survey we asked participants to honestly report whether they answered the questions in a ``sincere, conscious and engaged'' way, and whether they used ChatGPT or other LLMs\footnote{When we followed up with these participants via Prolific, four of the five excluded reported that they had misunderstood the questions and responded incorrectly. They were nonetheless excluded out of caution.}. This did not affect their compensation. 

After this, participants completed questionnaires for several psychological and UX measures (see next section). 
To check for careless responding, the survey also included three attention check questions. 
All questions in the survey, except age and gender, were mandatory. The median time to complete the survey was 18 minutes. For completing the survey, participants were compensated £9/h in accordance with Prolific's recommendation as of June 2023.

\subsection{Measures} \label{measures}
Unless noted otherwise, all measures were answered on a 7-point Likert scale, ranging from ``strongly disagree'' (1) to ``strongly agree'' (7).

\subsubsection{User Motivation Inventory (UMI)} \label{UMI}
To measure the six motivational regulations posited by OIT, we used the User Motivation Inventory (UMI) \cite{bruhlmann_measuring_2018}. This was chosen because it addresses all six motivational regulations, where other relevant scales \cite[e.g., ACTA][]{peters_designing_2018} do not. 
The set of scores for these regulations is referred to as a motivational profile \cite[e.g.,][]{bruhlmann_motivational_2020} (see \autoref{lpa_profiles}).  When a profile leans more toward the regulations on the right, this marks a relatively autonomous profile of motivation (see also \autoref{OIT_diagram}). When it leans more to the left, this marks a relatively low autonomy profile.
Cronbach's $\alpha$  scores indicated all sub-scales were reliable: $amotivation=0.91, external=0.88, introjection=0.79, identified=0.80, integrated=0.86, intrinsic=0.91$.

\subsubsection{Basic Need Satisfaction}
As noted in \autoref{bg_goodbad_ux}, SDT theorises that people have three basic psychological needs --- autonomy, competence and relatedness --- whose satisfaction supports growth, flourishing, and wellbeing \cite[][pp. 10-11]{ryan_self-determination_2017}.
We measure basic needs using Sheldon's need satisfaction scale \cite{sheldon_what_2001}, which measures satisfaction of these three basic needs alongside seven 
additional need constructs: \textit{Influence--popularity, Pleasure--stimulation, Security--control, Physical thriving--bodily, Self-actualising--meaning, Self-esteem--self-respect, Money--luxury}.

This scale has been commonly used in previous UX studies \cite[e.g.,][]{tuch_analyzing_2013, hassenzahl_needs_2010,mekler_momentary_2016} to provide additional context on dimensions of need satisfaction. Often researchers drop some items from the scale, e.g., addressing only the 3 basic needs \cite[e.g.,][]{bruhlmann_measuring_2018}, or dropping one or more of the constructs considered less relevant to the study at hand.
As in previous UX work \cite[e.g.,][]{tuch_analyzing_2013, hassenzahl_needs_2010, mekler_momentary_2016}, we drop the items for money-luxury and physical thriving which feel unrelated to typical technology interaction scenarios. This leaves eight needs in total. Of these we formed hypotheses about autonomy, competence, and relatedness. The remaining 5 needs are retained for exploratory analysis only: to shed further light on the results, inform future research directions, and to relate results to prior UX research.

Cronbach's $\alpha$ scores 
indicated satisfactory levels of reliability for all need sub-scales except Security--control, which fell slightly below the typical benchmark of 0.7 (i.e., $Autonomy=0.798, competence=0.88, relatedness=0.89$, $self-actualising-meaning=0.86, pleasure=0.77, security=0.66, self-esteem=0.83, popularity=0.85$).


\subsubsection{Attribution}
We expected that more autonomous
motivation might relate to a greater sense of responsibility for outcomes, since more autonomous forms of motivation are associated with the internalisation of motives, and internal perceived locus of control \cite{sheldon_becoming_2014}. 
We asked participants to answer three questions on 7-point Likert scales, concerning the degree to which they felt 1) they were responsible for outcomes of interactions with the technology (i.e., self-attribution), vs 2) the technology or 3) other factors being responsible. The latter two were intended to provide context to interpret the self-attribution item, but during analysis we did not find these informative. As such only the  self-attribution question is discussed. Interested readers may refer to the data shared in the \href{https://osf.io/eny6v/}{OSF repository}. 

\subsubsection{Usability Metric for User Experience (UMUX)}
We used the four-item UMUX scale \cite{finstad_usability_2010} to measure perceived usability and understand how product qualities relate to motivation.
The UMUX is shorter than other candidate measures for usability, and normative data is available to ground its interpretation \cite{hodrien_review_2021}.  It comprises four questions addressing the capabilities of the system, the level of frustration in use, the ease of use, and the need to correct system behaviour.
Cronbach's $\alpha$  for the UMUX (0.76) indicated good reliability.

\subsubsection{Positive and Negative Affect Schedule (PANAS)}
We used the short-form 20-item PANAS-C scale to measure positive and negative affect \cite{thompson_development_2007}, which has seen frequent use in UX studies \cite{tuch_analyzing_2013, hassenzahl_needs_2010,mekler_momentary_2016}. The PANAS has been shown to be stable over time, making it suitable for measuring both episodic and longer-term affect \cite{thompson_development_2007}. 
Items were answered on a 5-point Likert scale from "very slightly or not at all" to "extremely". To compute an overall affect score, means are taken for positive and negative items, and the negative score subtracted from the positive, leading to an overall scale of -5 to 5.
Cronbach's $\alpha$  for the positive scale was $0.89$, and for the negative scale $0.86$, indicating good reliability.

\subsection{Analysis of Open Questions}

To make our analysis sensitive to relevant factors we did not define a code book ahead of analysis. Instead, we took a bottom-up, or ``inductive'' approach, avoiding a priori assumptions. This was guided by descriptions of good practice in Thematic Analysis (TA) as described by Braun and Clarke \cite{braun_using_2006, braun_successful_2013, braun_one_2020}. 
All coding was completed by the first author, in NVIVO. Only responses to the open questions were exported to NVIVO, to avoid potential biases from knowing which 
experimental condition or profile 
participants were assigned to.  
The first and second author met periodically at each stage of the coding process to review sample codes and themes, as well as check their sense and consistency.

We first began by reading all responses to familiarise ourselves with the data, before generating codes. We 
coded any features which seemed relevant to the user's relationship to the technology. In practice this included, for example, the meaning and value of the technology, the context of use, the style of use, and evaluations of the technology.
A second pass of coding was then completed focusing on consistency of coding: grouping equivalent codes, and sense-checking the relation of extracts to codes.
Finally, we grouped these codes: first, into higher level codes, then into sub-themes and themes. To serve the goals of the study, we aimed to identify a small number of high-level themes, each of which retained a relatively high number of sub-themes. This facilitated granular comparison across motivational profiles. Specifically, we identified four high-level themes (Purpose; Qualities of the technology; Being good or bad for people; and Values, identity and emotional connection). 
These are described in \autoref{Results}.

Finally, we counted the percentage of participants in each profile whose extracts fell under each theme, subtheme, and high-level code. Rather than being interpreted as precise quantities, these percentages were used as a broader heuristic to identify differences across profiles. It is important to note that this final stage of analysis diverges significantly from reflexive TA practice. It de-emphasises reflexive TA's focus on analyst-subjectivity, but allows us to identify patterns that differ between motivational profiles, supporting interpretation. In this respect our approach is similar to content analysis, which in its qualitative form often also involves the inductive generation of a codebook \cite{elo_qualitative_2008}.
As such, while the inductive stage of analysis was guided by reflexive TA practices, we do not present the overall process, or results, as TA.
Full details of coding are included as supplementary materials\footnote{Refer to \url{https://osf.io/eny6v/} for supplementary material.}.

\section{Identifying Motivational Profiles}
The UMI measures the degree to which participants' technology use is characterised by six distinct motivational regulations \cite{bruhlmann_measuring_2018}. We used Latent Profile Analysis (LPA) to identify sub-groups of participants based on their similarity across these six motivational regulations.  We refer to these sub-groups as ``motivational profiles'' as in 
\cite[e.g.,][]{bruhlmann_motivational_2020,gustafsson_motivational_2018}.

We first conducted a confirmatory factor analysis on the UMI scores. This served to confirm the validity of the measurement model, and allowed us to use factor scores in the LPA analysis. This was followed by the LPA analysis itself, which identified the motivational profiles. 
Finally, we related the resulting profiles to our UX measures, 
using the Bolck, Croon, and Hagenaars (BCH) method \cite{bauer_primer_2021}, 
which is considered best practice for such analysis \cite{bauer_primer_2021}. 
Individual steps in this analysis are described in detail below.



\label{lpa_descriptive}

\paragraph{Confirmatory Factor Analysis (CFA)}
Previous work has argued that it is ``arguably optimal'' \cite[][p.~10]{bauer_primer_2021} to conduct LPA on factor scores (e.g., resulting from a CFA), rather than raw measure scores. This ensures that the indicators used are truly continuous (as required by LPA) and that measurement error has been corrected for \cite{bauer_primer_2021}. 
CFA is also an important first step in more advanced modelling since it tests the assumptions of the measurement model (here,  the six-factor model of the UMI).
As such, we specified a six-factor measurement model in which each item in the UMI loaded onto its designated factor. A requirement of the BCH approach is that factor scores are scaled to have meaningful mean structure \cite{bauer_primer_2021}. We ensured this via \textit{effects coding} --- constraining the model such that factor loadings average to 1 and indicator intercepts sum to 0 \cite{little_non-arbitrary_2006}. 
Since we did not observe multivariate normality in our results, we used a robust maximum likelihood estimator (the MLR method in the Lavaan package \cite{lavaan2012}), and evaluated fit statistics scaled by the Yan-Bentler correction.
Results of CFA indicated that the six factor model was supported by the data [$\chi^2 = 290.23, p < 0.01, \chi^2/df = 2.41, CFI = 0.965, SRMR = 0.059, RMSEA = 0.053, PCLOSE = 0.144$].


 \begin{figure*}[p]
  \includegraphics[width=0.95\linewidth]{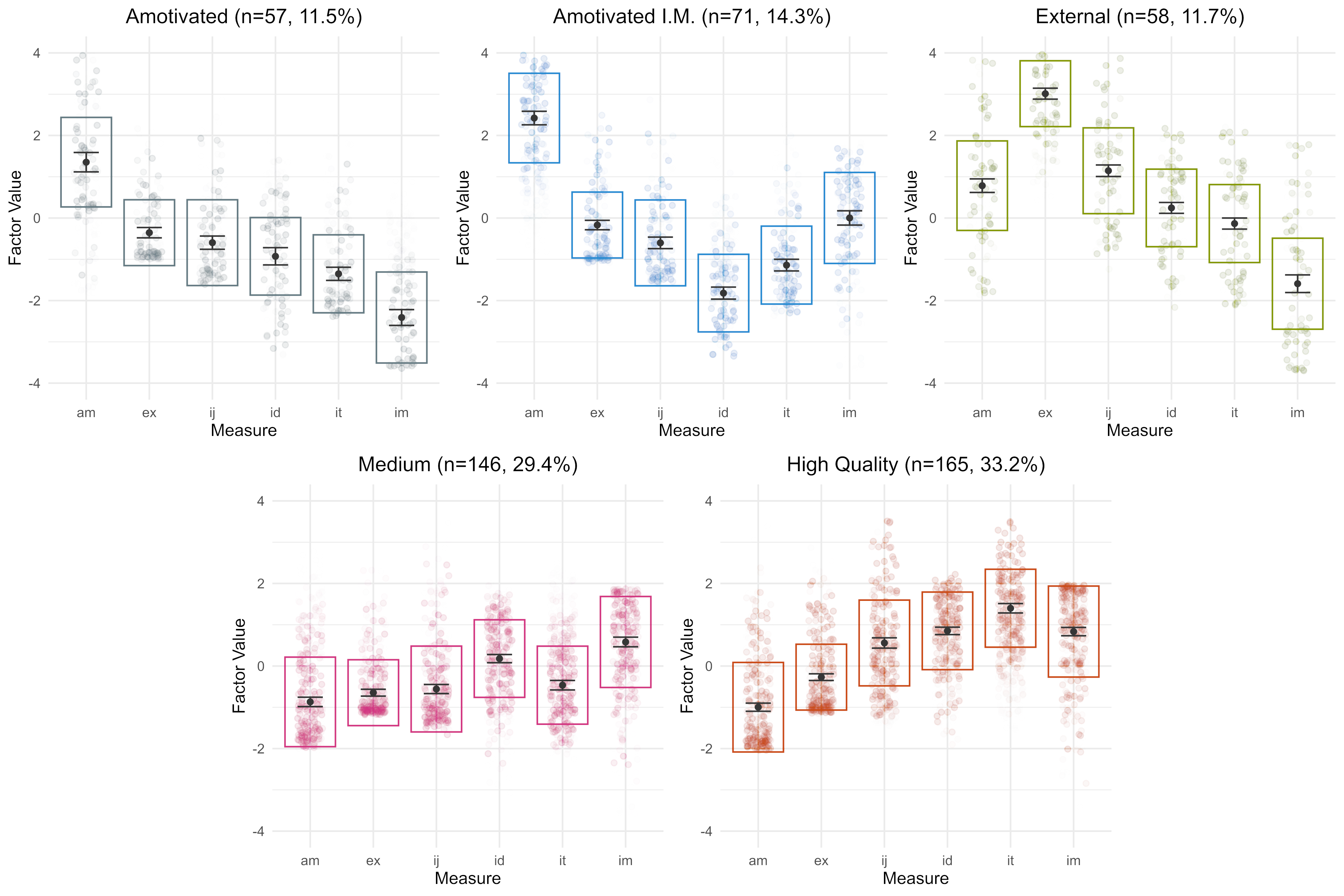}  
  \vspace{-1em}
  \caption{The five motivational profiles identified via latent profile analysis.  Higher scores on constructs to the left indicate a less autonomous motivational profile, and higher scores on constructs to the right indicate a more autonomous profile. The black points indicate the mean, error bars show standard errors, boxes show standard deviations. Profile membership in LPA is probabilistic, and membership probability is indicated here by opacity (range:$(0.82,0.95)$).  am = amotivated, ex = external regulation, ij = introjected regulation, id = identified regulation, it = integrated regulation, im = intrinsic motivation. 
}
  \label{lpa_profiles}
 \end{figure*}


\begin{table*}[p]
\caption{\label{tab_profiles_high_level}High-level summary of the profiles focusing on the most salient features. For more detail see results.}
\centering
\resizebox{0.97\linewidth}{!}{
\begin{tabular}{|p{0.12\textwidth}|p{0.22\textwidth}|p{0.55\textwidth}|p{0.24\textwidth}|}
\hline
\textbf{Profile} & \textbf{UX Measures} & \textbf{Kind of   engagement and attitude} & \textbf{Technologies} 
\\ \hline
\rowcolor[HTML]{FFFFFF} 
\cellcolor[HTML]{B0BCC0}Amotivated    & Very   low need satisfaction, affect and usability                                                           & Users   have poor perception of the technology, and often did not chose to use it   themselves. Often users felt the technology clashed with their values and was   bad for them and others.                                                                                                       & A   broad mix. Most prominent:  $\sim$25\% social   media.                                              \\ \hline
\rowcolor[HTML]{FFFFFF} 
\cellcolor[HTML]{9BCCED}Amotivated-Intrinsic  (Amotivated I.M.) & Low   need satisfaction with the exception of pleasure. Low affect. Moderate   usability.                    & Low-intentionality  and sometimes compulsive engagement.   Technology often clashes with users' values, and is considered bad for   them.                                                                                                                                                          & Mostly   social media apps and entertainment technologies.                                              \\ \hline
\rowcolor[HTML]{FFFFFF} 
\cellcolor[HTML]{E1E6C0}External      & Low   need satisfaction with the exception of popularity, security, relatedness.   Low affect and usability. & Poor   perception of the technology, which is often not self-chosen. Users sometimes   connect tech to personal insecurities. Otherwise they often feel it has   little meaning beyond the pragmatic.                                                                                              & Somewhat   mixed, Most prominent: $\sim$20\% each for social media, messaging, workplace   technologies \\ \hline
\rowcolor[HTML]{FFFFFF} 
\cellcolor[HTML]{E789B4}Medium        & Moderate   need satisfaction, but low relatedness. Moderate-to-high affect. High   usability                 & Often   used to pursue hobbies interests and passions, or for self-growth, but users   do not necessarily feel a personal connection or deep meaning. Users   generally descibe  the technology as   having positive qualities.                                                                    & A   broad mix. Most prominent:  $\sim$25\% social   media. $\sim$20\% streaming media.                  \\ \hline
\rowcolor[HTML]{FFFFFF} 
\cellcolor[HTML]{EBB39D}High Quality  & High   satisfaction of all needs. High affect, High usability.                                               & Conscious,   self driven engagement, often focused on hobbies interests and passions. More   likely than other groups to identify with the technology or attribute meaning   to it. Users commonly have a positive view of the technology itself, and even   feel the technology is good for them. & A   broad mix. Most prominent:  $\sim$25\% social   media. \\                        \hline
  
\end{tabular}
}
\end{table*}


\paragraph{Latent Profile Analysis (LPA)}
LPA allows the identification of sub-groups of participants who share a similar profile across a set of variables of interest. 
LPA uses mixture modelling techniques to identify clusters of observations whose values are similar across the variables of interest, adapting the approach of latent class analysis to continuous data \cite{bauer_primer_2021}. While LPA is a relatively novel technique in UX research, it is well established in psychology \cite{bauer_primer_2021, robertson_mixture_2016}, and particularly in SDT research \cite[e.g.,][]{wang_exploring_2017, gustafsson_motivational_2018, rouse_interplay_2020, bruhlmann_motivational_2020}.
We used the tidy-sem library in R \cite{van_lissa_recommended_2023} to apply the approach to identify distinct \textit{motivational profiles} in our sample: clusters of participants who are similar across the six motivational regulations measured by the UMI.

We applied LPA to the factor scores derived from the CFA, as this helps correct for measurement error \cite{bauer_primer_2021}. As is the norm with LPA \cite{bauer_primer_2021}, the number of profiles was determined empirically by evaluating a number of different models. 
In our pilot analysis data had best supported a 4-profile model \cite[aligning with results in][]{bruhlmann_motivational_2020}, and larger profile counts had resulted in high numbers of parameters per participant, risking overfitting. 
As such, we investigated models with profile counts ranging from 1 to 6.
In LPA
it is possible to enforce more-or-less strict modelling constraints, concerning indicator variances and the covariance matrix, and there is a lack of consensus about the value of different approaches \cite[see][]{peugh_modeling_2013, bauer_primer_2021, van_lissa_recommended_2023}. 
As such, we followed prior recommendations \cite{bauer_primer_2021} by testing models for all profile-counts, both with and without constraints on variance and covariance. 

The final model was selected based on a combination of model fit statistics and theoretical considerations \cite{bauer_primer_2021, howard2016motivation}.
We first eliminated models which failed to converge (all free covariance models). To avoid risk of overfitting, we eliminated models where the number of participants in the smallest class was less than five times the number of parameters. To support adequate power in later hypothesis testing, we also eliminated models where the smallest profile held less than 10\% of the sample ($\sim50$ participants).
For the remaining models, we evaluated model fit using the sample-size adjusted Bayesian Information Criterion (saBIC\footnote{saBIC is found to perform better than BIC in evaluating mixture models \cite{sen_comparison_2017}.}), while also taking into account the Log Likelihood (LL).
Optimal values of LL and saBIC were found in the 5-profile model estimated with equal variance, equal covariance. Minimum classification probability (i.e., of a user in a profile) was acceptable (0.82).

On the basis of these results, we deemed the 5-profile, equal variance, equal covariance model to be optimal\footnote{The full decision process and all data is recorded in the \href{https://osf.io/eny6v/}{OSF repository.}}. This model is shown in \autoref{lpa_profiles}, illustrating the distribution of scores for the six 
motivational regulations for each of the profiles. Profiles were calculated on factor scores for each construct (a weighted sum of the items in a given construct, weighted by factor loadings), and a score of zero represents 
the centre value for the distribution of the factor scores for each construct. For this reason, zero on the factor scores may not correspond precisely to the reported mean in \autoref{tab:umi_descriptive}, though they can be expected to be close.

\subsection{Relating Profiles to Outcome Variables} \label{sec:relating_the_profiles}
When linking LPA profiles to outcome variables, previous work has often taken a ``naive'' \cite[][p. 341]{bakk_relating_2021} approach, where 
participants are assigned to profiles, and then mean outcome variables are compared between profiles, e.g., using ANOVA. This approach is misleading as LPA does not directly assign data points to profiles, but rather assigns probabilities, which ANOVA does not take into account \cite{bauer_primer_2021, bakk_relating_2021}. 
Instead, we opted for the Bolck–Croon–Hagenaars (BCH) method for linking profiles to outcome variables. 
This was 
followed by hypothesis testing with 
likelihood ratio and Wald tests \cite{bauer_primer_2021, bakk_relating_2021}. This approach takes into account classification uncertainty by linking profiles to outcomes via weighted estimation of a linear regression model. It also avoids normality assumptions and is insensitive to heteroscedasticity \cite{bakk_relating_2021}.


\section{Results}
\label{Results}

Overall, the most commonly selected technology
was social media (31\%) (see also \autoref{tab:profile_technologies} and \autoref{tab:tech_list}). Other technologies included  
streaming media (9\%), media authoring apps (6\%), voice assistants (4\%) and AI Chatbots (3\%).  
The median usage frequency 
was 2-3 times per day. Most participants (n = 201) used the technology at least 4 times per day, and only 15 used the technology less than once per week. 

In the following, we first report high-level results for the overall sample and the positive and negative conditions. We then report results for motivational profiles and the qualitative analysis.

\begin{table}

\caption{\label{tab:profile_technologies}Ten most frequent technology categories overall, per profile and per condition. 
}
\vspace{-1em}
\centering
\resizebox{\linewidth}{!}{
\begin{tabular}[t]{>{}l|r>{}r|rrrr>{}r|r}
\toprule
Technology Category & Positive & Negative & High 
Quality & Medium & External & Amotivated
 IM & Amotivated & Total\\
\midrule
\cellcolor{gray!6}{social media} & \cellcolor{gray!6}{26\% (67)} & \cellcolor{gray!6}{36\% (87)} & \cellcolor{gray!6}{25\% (41)} & \cellcolor{gray!6}{25\% (36)} & \cellcolor{gray!6}{21\% (12)} & \cellcolor{gray!6}{73\% (52)} & \cellcolor{gray!6}{23\% (13)} & \cellcolor{gray!6}{154}\\
streaming media & 12\% (30) & 6\% (14) & 8\% (13) & 18\% (26) & 0\% (0) & 6\% (4) & 2\% (1) & 44\\
\cellcolor{gray!6}{messaging and Email} & \cellcolor{gray!6}{7\% (17)} & \cellcolor{gray!6}{7\% (17)} & \cellcolor{gray!6}{7\% (11)} & \cellcolor{gray!6}{5\% (8)} & \cellcolor{gray!6}{17\% (10)} & \cellcolor{gray!6}{3\% (2)} & \cellcolor{gray!6}{5\% (3)} & \cellcolor{gray!6}{34}\\
creativity, media, writing & 9\% (22) & 4\% (10) & 8\% (13) & 8\% (11) & 5\% (3) & 1\% (1) & 7\% (4) & 32\\
\cellcolor{gray!6}{computers, phones, tablets} & \cellcolor{gray!6}{5\% (12)} & \cellcolor{gray!6}{4\% (9)} & \cellcolor{gray!6}{5\% (9)} & \cellcolor{gray!6}{5\% (8)} & \cellcolor{gray!6}{3\% (2)} & \cellcolor{gray!6}{0\% (0)} & \cellcolor{gray!6}{4\% (2)} & \cellcolor{gray!6}{21}\\
voice assistants & 4\% (10) & 5\% (11) & 4\% (6) & 6\% (9) & 0\% (0) & 1\% (1) & 9\% (5) & 21\\
\cellcolor{gray!6}{finance} & \cellcolor{gray!6}{5\% (12)} & \cellcolor{gray!6}{3\% (8)} & \cellcolor{gray!6}{5\% (8)} & \cellcolor{gray!6}{5\% (7)} & \cellcolor{gray!6}{3\% (2)} & \cellcolor{gray!6}{0\% (0)} & \cellcolor{gray!6}{5\% (3)} & \cellcolor{gray!6}{20}\\
workplace technologies & 2\% (6) & 6\% (14) & 2\% (4) & 1\% (2) & 22\% (13) & 0\% (0) & 2\% (1) & 20\\
\cellcolor{gray!6}{gaming} & \cellcolor{gray!6}{4\% (11)} & \cellcolor{gray!6}{3\% (8)} & \cellcolor{gray!6}{2\% (4)} & \cellcolor{gray!6}{5\% (8)} & \cellcolor{gray!6}{3\% (2)} & \cellcolor{gray!6}{7\% (5)} & \cellcolor{gray!6}{0\% (0)} & \cellcolor{gray!6}{19}\\
fitness, health and wellbeing & 5\% (12) & 2\% (4) & 8\% (13) & 1\% (1) & 0\% (0) & 0\% (0) & 4\% (2) & 16\\
\bottomrule
\end{tabular}}
\end{table}


 \begin{figure}
\includegraphics[width=1\linewidth]{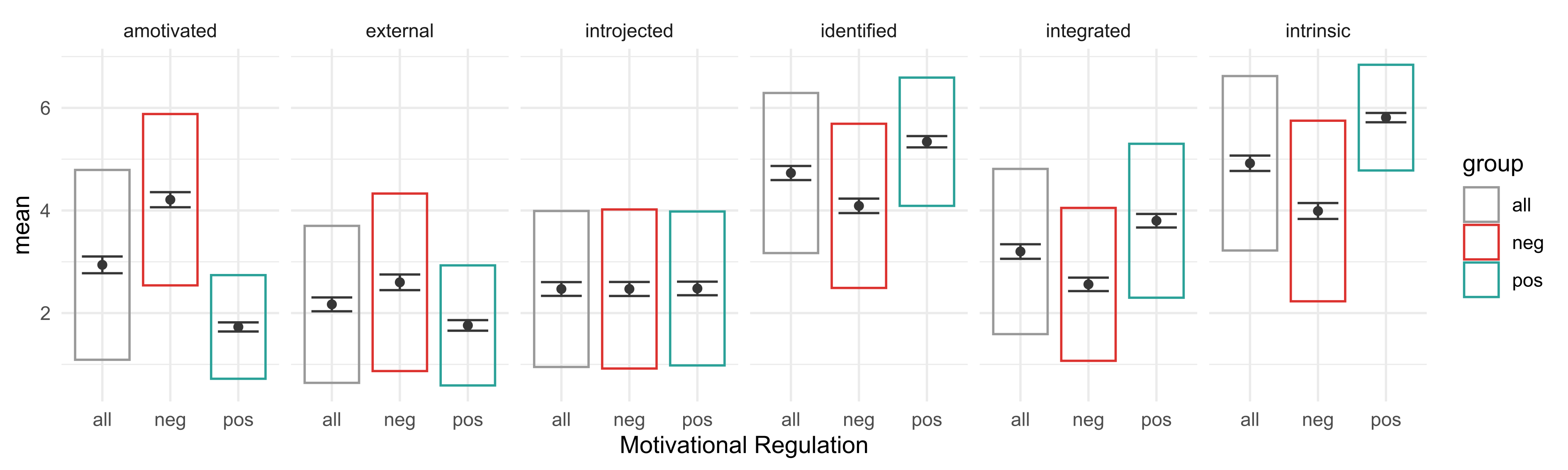}  
  \caption{Plots of the UMI data for the 6 motivational regulations, over the whole data set (all), and split by condition (negative, positive).  Data provided in \autoref{tab:umi_descriptive} in \autoref{appA}. Points show means, error bars show standard error, boxes show standard deviation}
  \label{plt:umi_conditions}
 \end{figure}

\subsection{Characteristics of Positive and Negative Technology Use}

\paragraph{Motivational Regulation}
Overall, participants displayed relatively autonomous patterns of motivation (see \autoref{plt:umi_conditions}). 
This leaning towards autonomous regulations was stronger in the positive condition, which had lower values for amotivation (M = 1.73) and extrinsic motivation (M = 1.76), and higher values for identified (M = 5.34), integrated (M = 3.8) and intrinsic (M = 5.81) regulations. 
In the negative condition 
amotivation scored highest. 
Mean scores for identified and intrinsic regulations in the negative condition are just above scale midpoint, indicating that these participants still felt their use of the selected technology 
was at least somewhat worthwhile and inherently satisfying.


\paragraph{Hypothesis Variables}
As listed in \autoref{tab:bns_var}, all three basic needs received higher scores in the positive condition than the negative condition, with no overlap in confidence intervals. 
Similarly, PANAS, UMUX, and Self-Attribution received higher scores in the positive condition than the negative condition (see \autoref{tab:other_hyp_var}). 
In both positive and negative conditions the mean scores for the PANAS were positive (M = 1.96 and M = 0.61, respectively). 
Finally, means for self-attribution were slightly higher in the positive condition (M = 5.66 vs. M = 5.13), though both were comfortably above the midpoint, indicating that users in both conditions tended to feel a sense of responsibility for the outcomes experienced.  
\begin{table}

\caption{\label{tab:bns_var}Basic Need Satisfaction variables per profile. M = mean, CI = confidence interval (lower, upper), SD = standard deviation.}
\vspace{-1em}
\centering
\resizebox{\linewidth}{!}{
\begin{tabular}[t]{>{\raggedright\arraybackslash}p{9em}rrccccccccc}
\toprule
\multicolumn{3}{c}{ } & \multicolumn{3}{c}{Autonomy} & \multicolumn{3}{c}{Competence} & \multicolumn{3}{c}{Relatedness} \\
\cmidrule(l{3pt}r{3pt}){4-6} \cmidrule(l{3pt}r{3pt}){7-9} \cmidrule(l{3pt}r{3pt}){10-12}
 & Participants & \% Positive & M & CI & SD & M & CI & SD & M & CI & SD\\
\midrule
\textbf{\cellcolor{gray!6}{Mean Positive}} & \cellcolor{gray!6}{254 (51\%)} & \cellcolor{gray!6}{100.0} & \cellcolor{gray!6}{4.75} & \cellcolor{gray!6}{{}[4.59, 4.91]} & \cellcolor{gray!6}{1.32} & \cellcolor{gray!6}{4.31} & \cellcolor{gray!6}{{}[4.1, 4.53]} & \cellcolor{gray!6}{1.73} & \cellcolor{gray!6}{3.61} & \cellcolor{gray!6}{{}[3.38, 3.84]} & \cellcolor{gray!6}{1.85}\\
\textbf{Mean Negative} & 243 (49\%) & 0.0 & 3.38 & {}[3.19, 3.58] & 1.53 & 3.25 & {}[3.04, 3.46] & 1.64 & 2.93 & {}[2.71, 3.15] & 1.74\\
\midrule
\textbf{\cellcolor{gray!6}{High Quality}} & \cellcolor{gray!6}{165 (33.2\%)} & \cellcolor{gray!6}{74.5} & \cellcolor{gray!6}{5.47} & \cellcolor{gray!6}{{}[5.34, 5.6]} & \cellcolor{gray!6}{0.87} & \cellcolor{gray!6}{5.07} & \cellcolor{gray!6}{{}[4.86, 5.28]} & \cellcolor{gray!6}{1.36} & \cellcolor{gray!6}{4.09} & \cellcolor{gray!6}{{}[3.82, 4.35]} & \cellcolor{gray!6}{1.74}\\
\textbf{Medium} & 146 (29.4\%) & 76.0 & 3.98 & {}[3.79, 4.18] & 1.21 & 3.54 & {}[3.28, 3.81] & 1.65 & 2.97 & {}[2.69, 3.25] & 1.74\\
\textbf{\cellcolor{gray!6}{External}} & \cellcolor{gray!6}{58 (11.7\%)} & \cellcolor{gray!6}{19.0} & \cellcolor{gray!6}{3.26} & \cellcolor{gray!6}{{}[2.85, 3.66]} & \cellcolor{gray!6}{1.57} & \cellcolor{gray!6}{3.86} & \cellcolor{gray!6}{{}[3.47, 4.25]} & \cellcolor{gray!6}{1.50} & \cellcolor{gray!6}{3.80} & \cellcolor{gray!6}{{}[3.28, 4.32]} & \cellcolor{gray!6}{1.99}\\
\textbf{Amotivated I.M.} & 71 (14.3\%) & 9.9 & 3.07 & {}[2.78, 3.35] & 1.23 & 2.08 & {}[1.86, 2.3] & 0.94 & 2.72 & {}[2.39, 3.05] & 1.42\\
\textbf{\cellcolor{gray!6}{Amotivated}} & \cellcolor{gray!6}{57 (11.5\%)} & \cellcolor{gray!6}{3.5} & \cellcolor{gray!6}{2.34} & \cellcolor{gray!6}{{}[2.09, 2.59]} & \cellcolor{gray!6}{0.95} & \cellcolor{gray!6}{2.79} & \cellcolor{gray!6}{{}[2.43, 3.15]} & \cellcolor{gray!6}{1.36} & \cellcolor{gray!6}{1.84} & \cellcolor{gray!6}{{}[1.54, 2.14]} & \cellcolor{gray!6}{1.13}\\
\bottomrule
\end{tabular}}
\end{table}

\begin{table}

\caption{\label{tab:other_hyp_var}Other hypothesis variables per profile. M = mean, CI = confidence interval (lower, upper), SD = standard deviation.}
\vspace{-1em}
\centering
\resizebox{\linewidth}{!}{
\begin{tabular}[t]{>{\raggedright\arraybackslash}p{9em}rrccccccccc}
\toprule
\multicolumn{3}{c}{ } & \multicolumn{3}{c}{PANAS} & \multicolumn{3}{c}{UMUX} & \multicolumn{3}{c}{Self-Attribution} \\
\cmidrule(l{3pt}r{3pt}){4-6} \cmidrule(l{3pt}r{3pt}){7-9} \cmidrule(l{3pt}r{3pt}){10-12}
 & Participants & \% Positive & M & CI & SD & M & CI & SD & M & CI & SD\\
\midrule
\textbf{\cellcolor{gray!6}{Mean Positive}} & \cellcolor{gray!6}{254 (51\%)} & \cellcolor{gray!6}{100.0} & \cellcolor{gray!6}{1.96} & \cellcolor{gray!6}{{}[1.86, 2.07]} & \cellcolor{gray!6}{0.85} & \cellcolor{gray!6}{6.19} & \cellcolor{gray!6}{{}[6.09, 6.29]} & \cellcolor{gray!6}{0.81} & \cellcolor{gray!6}{5.66} & \cellcolor{gray!6}{{}[5.48, 5.84]} & \cellcolor{gray!6}{1.45}\\
\textbf{Mean Negative} & 243 (49\%) & 0.0 & 0.61 & {}[0.47, 0.74] & 1.08 & 4.54 & {}[4.34, 4.73] & 1.55 & 5.13 & {}[4.91, 5.35] & 1.73\\
\midrule
\textbf{\cellcolor{gray!6}{High Quality}} & \cellcolor{gray!6}{165 (33.2\%)} & \cellcolor{gray!6}{74.5} & \cellcolor{gray!6}{2.35} & \cellcolor{gray!6}{{}[2.24, 2.46]} & \cellcolor{gray!6}{0.72} & \cellcolor{gray!6}{6.01} & \cellcolor{gray!6}{{}[5.85, 6.17]} & \cellcolor{gray!6}{1.07} & \cellcolor{gray!6}{5.88} & \cellcolor{gray!6}{{}[5.71, 6.05]} & \cellcolor{gray!6}{1.11}\\
\textbf{Medium} & 146 (29.4\%) & 76.0 & 1.55 & {}[1.45, 1.66] & 0.65 & 5.97 & {}[5.81, 6.13] & 0.99 & 5.29 & {}[4.99, 5.58] & 1.82\\
\textbf{\cellcolor{gray!6}{External}} & \cellcolor{gray!6}{58 (11.7\%)} & \cellcolor{gray!6}{19.0} & \cellcolor{gray!6}{0.36} & \cellcolor{gray!6}{{}[0.07, 0.64]} & \cellcolor{gray!6}{1.08} & \cellcolor{gray!6}{4.03} & \cellcolor{gray!6}{{}[3.62, 4.45]} & \cellcolor{gray!6}{1.60} & \cellcolor{gray!6}{5.23} & \cellcolor{gray!6}{{}[4.79, 5.66]} & \cellcolor{gray!6}{1.68}\\
\textbf{Amotivated I.M.} & 71 (14.3\%) & 9.9 & 0.21 & {}[0.07, 0.36] & 0.62 & 5.52 & {}[5.31, 5.74] & 0.94 & 5.57 & {}[5.24, 5.9] & 1.43\\
\textbf{\cellcolor{gray!6}{Amotivated}} & \cellcolor{gray!6}{57 (11.5\%)} & \cellcolor{gray!6}{3.5} & \cellcolor{gray!6}{-0.14} & \cellcolor{gray!6}{{}[-0.22, -0.05]} & \cellcolor{gray!6}{0.32} & \cellcolor{gray!6}{3.15} & \cellcolor{gray!6}{{}[2.9, 3.39]} & \cellcolor{gray!6}{0.94} & \cellcolor{gray!6}{4.23} & \cellcolor{gray!6}{{}[3.76, 4.7]} & \cellcolor{gray!6}{1.79}\\
\bottomrule
\end{tabular}}
\end{table}

\subsection{Statistical Comparison of the Profiles} \label{sec:statistical_comparison}
Statistical testing supported the rejection of the null hypothesis for H1, indicating significant differences in UX outcomes for all pairwise comparisons between profiles (see \autoref{tab:wald_classes} and \autoref{tab:hypothesis_results} in \autoref{appA}).
First, a likelihood ratio test indicated that the UX variables differed significantly between profiles $\chi^2(24)=1400.1, p < 0.0001$. Further, all pairwise comparisons of profiles also indicated significant differences in the UX measures (see \autoref{tab:wald_classes}).
As such, we used Wald tests to make pairwise comparisons between profiles for individual hypothesis variables. We corrected p-values for the 60 resulting hypotheses using the Holm-Bonferroni correction (via the \textit{adjust} function in R). 
After adjustment 48 comparisons gave significant results at $p < 0.05$, and 33 of these were significant at $p < 0.001$.  Significance levels of comparisons are indicated on \autoref{bch_hyp_plot} and \ref{bch_hyp_plot_2}. Full results for all tests are tabulated in \autoref{tab:hypothesis_results}, in \autoref{appA}.

\subsection{Characteristics of the Motivational Profiles} \label{sec:profile_descriptions}
In this section we highlight the characteristics of each of the five profiles identified via latent profile analysis. A high-level summary of these characteristics is provided in \autoref{tab_profiles_high_level}.

\subsubsection{Motivational regulations}

As noted in \autoref{sec:bg_OIT}, motivational profiles can be ordered in terms of their relative autonomy, i.e., their leaning toward more or less autonomous motivational regulations. Of all the profiles, the \textit{High Quality} profile is most autonomous in this sense, 
with particularly high scores for integrated regulation, the most autonomous form of extrinsic motivation.
Next is the \textit{Medium} profile where, again, autonomous regulations are most prominent. This profile scores similarly to the High Quality profile for intrinsic motivation, but scores
lower on all other regulations except amotivation, 
indicating a lower \textit{quantity} of motivation to engage with the technology in addition to the slightly less autonomous \textit{quality} of motivation. 

The next profile, \textit{External}, marks a larger qualitative shift, dominated by (
controlled) external and introjected regulations. Amotivation in this profile is higher than either \textit{High Quality} or \textit{Medium} profiles. 

The two least autonomously regulated profiles were both dominated by amotivation --- indicating a lack of intentionality and conscious valuing. 
The \textit{Amotivated} profile shows low scores on all regulations except amotivation, and in particular very little 
autonomous regulation. 
However, we also observed a profile with a more scooped motivational structure, peaking in amotivation and intrinsic motivation --- the \textit{Amotivated-Intrinsic} profile. This profile showed even higher levels of amotivation, while extrinsic regulations (and especially autonomous extrinsic regulations) were low: indicating little conscious valuing of the activity. At the same time, intrinsic motivation was almost as high as in the \textit{Medium} profile, indicating that 
motivation to engage with the technology was dominated by immediate interest and enjoyment.

\subsubsection{Technologies}
A full breakdown of technologies per profile is found in \autoref{tab:profile_technologies}. In all profiles the most common category was social media (20-25\%). 
However, only one profile was strongly dominated by a particular kinds technology: in the \textit{Amotivated-Intrinsic} profile 73\% of participants reported on social media technologies, with another 17\% reporting on entertainment technologies. 
The next most homogenous sample was the \textit{External} profile where workplace technologies (22\%), and messaging and email (17\%) were relatively prominent, within an otherwise diverse sample.
In the other profiles the technology types addressed were more diverse:
The \textit{High Quality}, \textit{Medium}, and \textit{Amotivated} profiles showed fairly similar distributions of technology types, though 
the \textit{High Quality} profile contained more accounts of fitness, health and wellbeing, 
the Medium profile contained more accounts of streaming media, 
and the Amotivated profile showed 
relatively high use of shopping apps. 

In general, a higher proportion of the participants in the more autonomously motivated profiles indicated that they felt good about the technologies they described (see \autoref{tab:bns_var}). However, even in the \textit{High Quality} and \textit{Medium} profiles around 25\% of participants reported on technologies they did not feel good about. In the \textit{External} profile only 19\% felt good about the technology, and this number was even lower for the \textit{Amotivated-Intrinsic} (9.9\%) and \textit{Amotivated} (3\%) profiles.


\begin{figure}[h]
\includegraphics[width=1\linewidth]{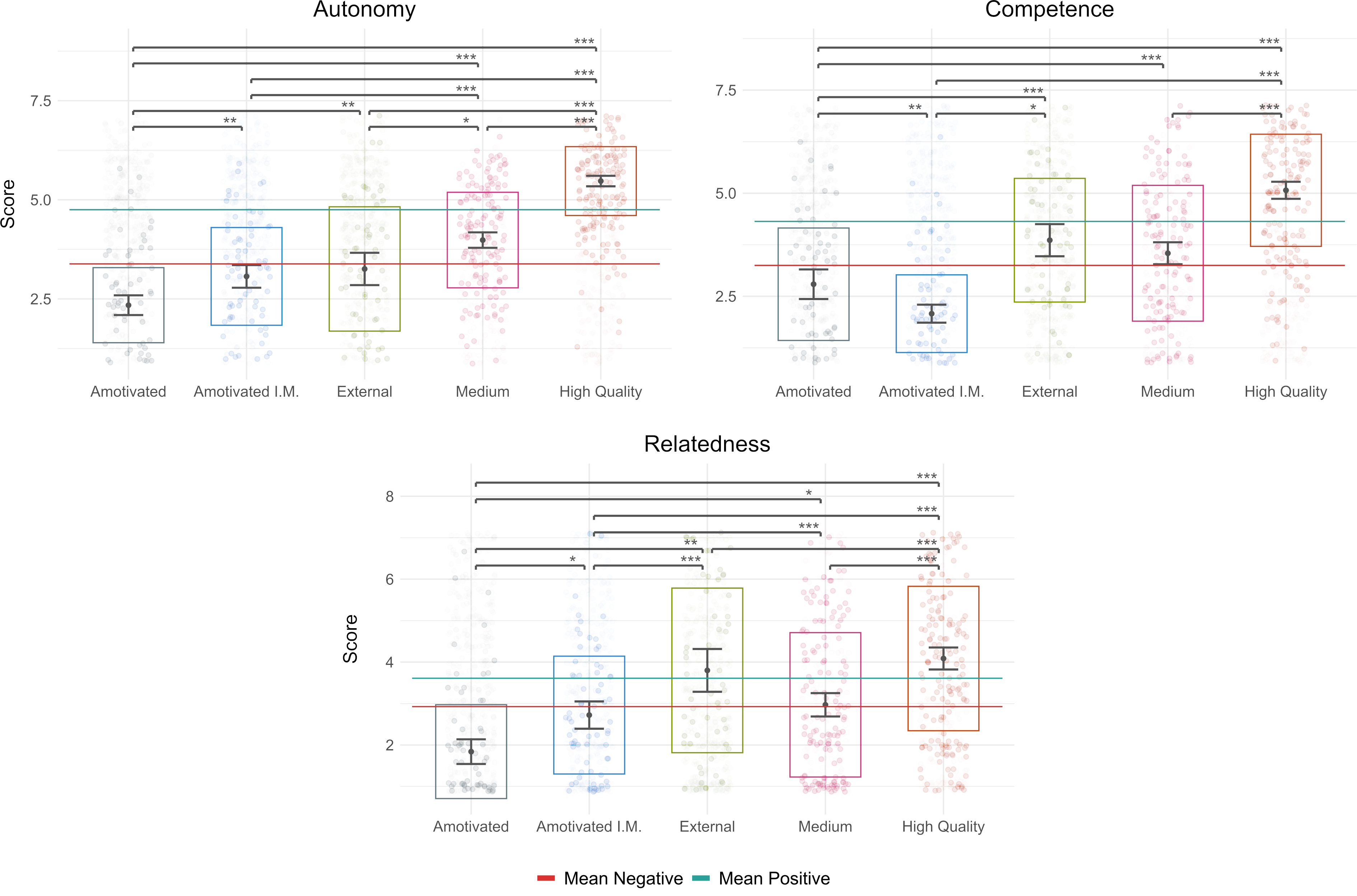}  
\vspace{-2em}
  \caption{Results for the three basic need measures. Black points indicate means, error bars show standard error, boxes show standard deviation. Green and red lines indicate indicates means for positive and negative' conditions respectively. Asterisks indicate Holm-Bonferroni adjusted p-values: *** $< 0.001$, ** $< 0.01$, * $< 0.05$. See \autoref{lpa_profiles} for more details supporting interpretation of the chart. Full tables of results, and plots for non-hypothesis variables are provided in \autoref{appA}.}
  \label{bch_hyp_plot}
 \end{figure}

\begin{figure}[ht]
\includegraphics[width=1\linewidth]{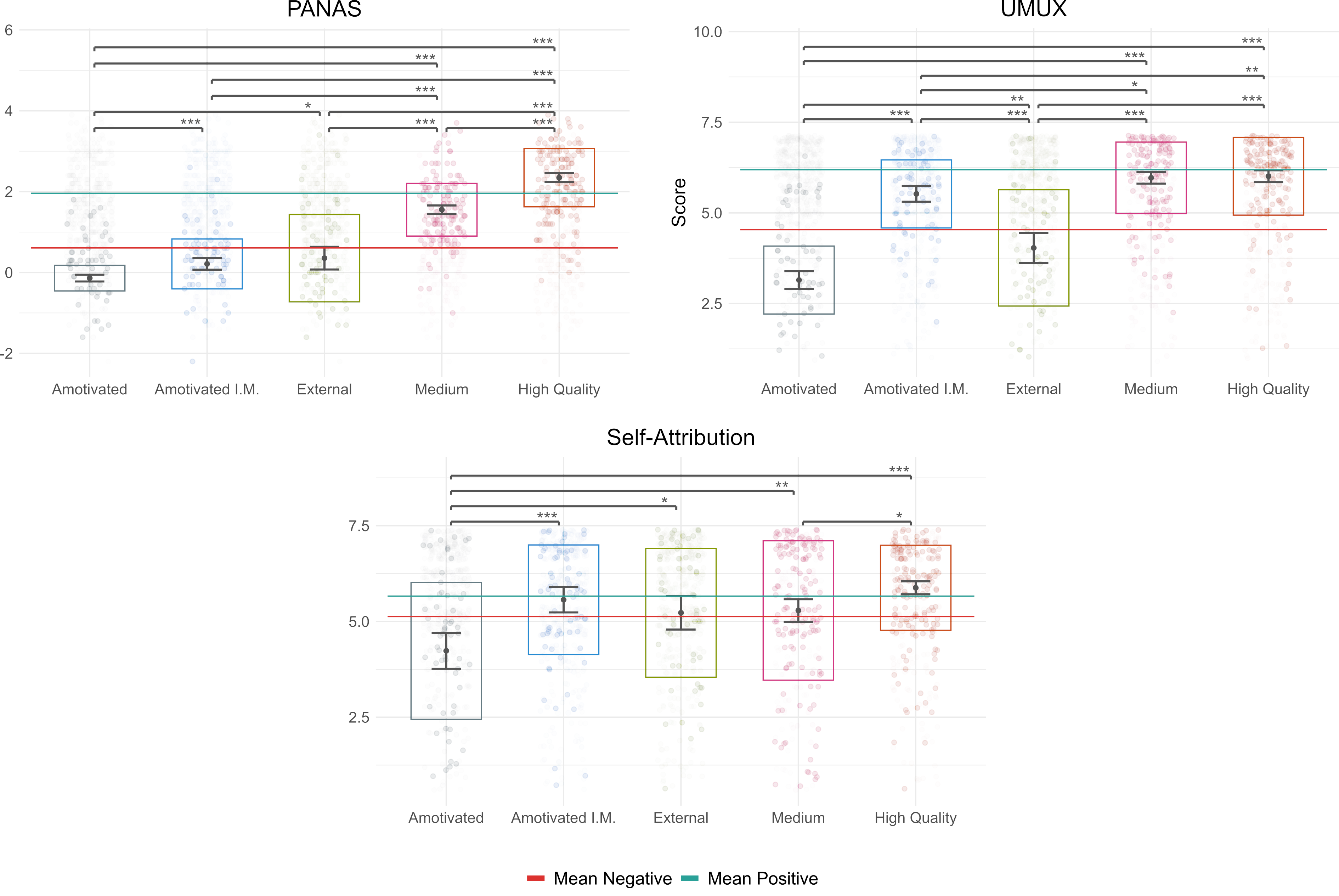}  
\vspace{-2em}
  \caption{Results for the PANAS (affect), UMUX (usability), and self-attribution measures. Black points indicate means, error bars show standard error, boxes show standard deviation. Green and red lines indicate indicates means for positive and negative' conditions respectively. Asterisks indicate Holm-Bonferroni adjusted p-values: *** $< 0.001$, ** $< 0.01$, * $< 0.05$. See \autoref{lpa_profiles} for more details supporting interpretation of the chart. Full tables of results, and plots for non-hypothesis variables are provided in \autoref{appA}.}
  \label{bch_hyp_plot_2}
 \end{figure}

\begin{figure}[ht]
  \includegraphics[width=1\linewidth]{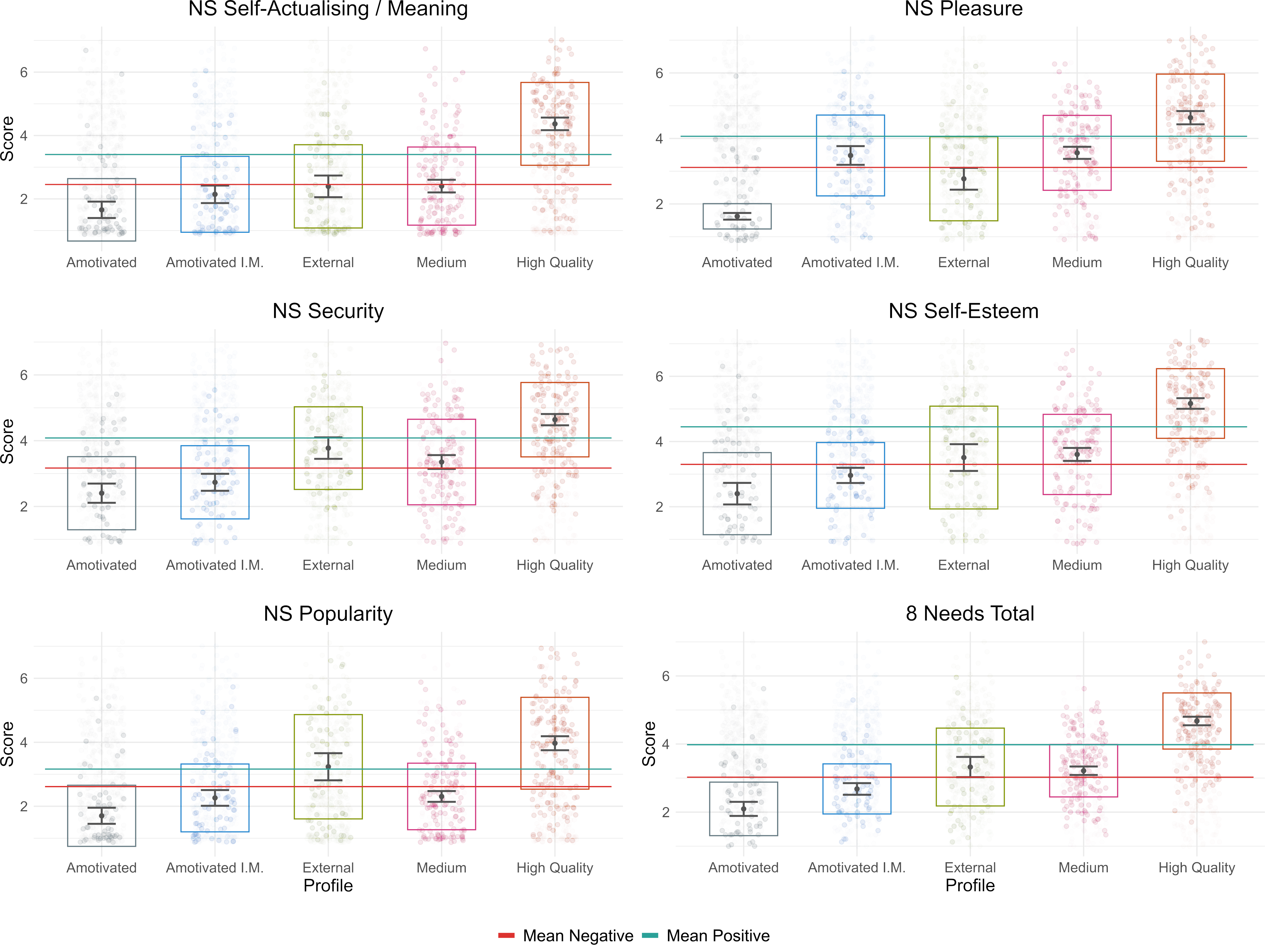}  
  \vspace{-2em}
  \caption{Results for the non-hypothesis variables. Black points indicate means, error bars show standard error, boxes show standard deviation. Green and red lines indicate indicates means for positive and negative conditions respectively. See \autoref{lpa_profiles} for more details supporting interpretation of the chart. }
  \label{bch_expl_plot}
 \end{figure}

\subsubsection{UX ratings}
As described in \autoref{sec:statistical_comparison}, we observed significant differences between profiles on all UX measures tested (see \autoref{bch_hyp_plot}, \autoref{bch_hyp_plot_2} and \autoref{bch_expl_plot}). Here we discuss results for individual measures.

\paragraph{Need Satisfaction.}
Need satisfaction is consistently higher in more autonomously motivated profiles. This pattern is most pronounced in the scores for autonomy, meaning and self-esteem.
Comparing adjacent profiles, the largest difference is 
found between the \textit{Medium} and \textit{High Quality} profiles. 
For the most autonomous profile --- \textit{High Quality} --- mean scores for all needs are comfortably above the respective mean scores, both overall and for the positive condition (see \autoref{tab:bns_var} and \autoref{bch_hyp_plot}). 
For the two least autonomous profiles --- \textit{Amotivated} and \textit{Amotivated-Intrinsic}  --- mean scores for all needs are comfortably below the respective mean scores for both the complete sample, and the negative condition. 

The \textit{External} profile represents one exception to this general pattern, scoring slightly above the more autonomous \textit{Medium} profile 
on relatedness, competence, popularity, and security  (albeit these differences are not statistically significant). 
Another notable exception 
is the \textit{Amotivated-Intrinsic} profile where pleasure scores are considerably higher than in the \textit{Amotivated} profile, and equivalent to those in the 
more autonomously motivated \textit{Medium} profile.

\paragraph{Affect} 
More autonomously motivated profiles scored consistently higher on affect, with pronounced differences even between adjacent profiles. \textit{External}, \textit{Amotivated-Intrinsic} and \textit{Amotivated} profiles, in contrast, featured fairly neutral affect scores. 

\paragraph{Usability} 
UMUX scores showed lower variation between profiles, and a less pronounced pattern overall.
\textit{High Quality} and \textit{Medium} profiles scored relatively high on usability, while 
the \textit{External} profile was close to the midpoint of the UMUX scale. 
Only the \textit{Amotivated} profile featured very low UMUX scores. 
In contrast, the \textit{Amotivated-Intrinsic} profile showed quite high UMUX scores --- only a little below 
the \textit{High Quality} profile.

\paragraph{Self-Attribution} 
In comparison to other metrics, self-attribution scores were relatively undifferentiated. 
Even the \textit{Amotivated-Intrinsic} profile was not significantly different from the \textit{High-Quality} profile. An exception to this trend was the \textit{Amotivated} profile which featured much lower self-attribution scores than all other profiles (see \autoref{bch_hyp_plot_2}).

\subsubsection{Free Text Answers}
Participants' written accounts contained 147 words on average (SD = 67), ranging from 25 to 621 words. Three participants wrote less than 50 words, and 85 wrote more than 200 words. 
From these accounts we identified five high-level themes, with nested sub-themes. Full coding is provided as supplementary material, and 
the most prominent themes are listed in 
\autoref{tab:themes}. Note that participant quotes are retained in their original spelling.

\paragraph{Purpose and Manner of Engagement}

Participants' accounts of their experiences differed substantially between profiles. Active and self-driven engagement was particularly common in the autonomous \textit{High Quality} (48\%) and \textit{Medium} profiles (38\%). 
More often than other profiles, these profiles described using the technology to pursue passions and interests ($\sim27\%$) or for self-growth and health ($\sim40\%$):
e.g., `\textit{`i  have learned painting through ticktok  and it has now became my new hobby. I  also use it to get more insights on spiritual teachings so i can advance in  my spiritual journey.''} (ID390).
Such active and self-driven engagement was less common in the \textit{Extrinsic} and \textit{Amotivated} profiles ($\sim25\%$), and least common in the \textit{Amotivated-Intrinsic} profile (8\%). 

Across profiles it was fairly common to use the technology for entertainment, and to connect with others. Entertainment was most commonly mentioned in the \textit{Amotivated-Intrinsic} (49\%), followed by the \textit{Medium} (34\%), \textit{High Quality} (28\%), \textit{Amotivated} (21\%), and \textit{External} (5\%) profiles. 
Connecting with others was most commonly mentioned in the \textit{External} profile (40\%). 

Participants in the \textit{Amotivated-Intrinsic} profile were by far most likely 
to describe their engagement in terms of compulsion and empty habits (63\% vs 20\% in the \textit{Amotivated} profile, and <10\% in other profiles), or in terms of escapism (32\% vs 14\% in the \textit{Medium} profile and < 10\% in other profiles). Participants in this profile were also more likely to describe passive consumption and lurking behaviour on social and streaming media (32\% vs <20\% in other profiles).
Finally, the \textit{External} (55\%) and \textit{Amotivated} (30\%) profiles were most likely to emphasise they did not engage with the technology by choice, 
e.g., ``I have to continue using it cause everyone else does'' (ID672). 

\paragraph{Values, Identity and Emotional Connection}
Participants often discussed emotional and value-based relationships with their chosen technologies.
Mentions of value \textit{disagreement} were more common than \textit{agreement} in all profiles except the \textit{High Quality} profile, and description of positive value agreement was relatively rare in general (in about 5\% of all descriptions). 
A few participants in the \textit{High Quality} (6\%) and \textit{Medium} (7\%) profiles described ways in which the technologies aligned with ethical values, or promoted fairness, sustainability and other positive values. 
Aside from explicitly relating the technology to shared values, participants also positively connected the technology to their own identities and values. In the \textit{High Quality} profile in particular some participants (19\%) prefaced positive evaluations of the technology with identity statements such as ``as a DJ'' (ID152), ``as a woman and as a mom'' (ID134), or connected the technology to professional and enthusiast identities such as ``content creator'' (ID523), ``marketing manager'' (ID233). 
Even where users in the High Quality profile described technologies they didn't feel good about (often due to unreliability or difficulty), the technology often connected to the self and values, e.g., supporting personally meaningful work (ID177) or giving ``ideas and motivation to elevate your life'' (ID357).

Mentions of value disagreement were most prevalent in the \textit{Amotivated-Intrinsic} profile (38\% of responses). This largely focused on poor perception of other users, and a sense of fakeness, alongside concerns with 
technology companies (e.g., \textit{`` It represents the aggresive capitalism around me which goes against my values.''} (ID295)). 
Such comments were less common in other profiles (13\% average for the whole sample) and particularly rare in the \textit{High Quality} profile (4\%).   
Users in the \textit{Amotivated-Intrinsic} profile were also more likely to discuss personal value conflicts than other profiles (17\% vs <5\% in other profiles): for example, ``Its a weird hate/love relationship. Instagram can be very inspiring and motivating at times [...]. Silmultanesly it will hold me back from achieving my dreams because I wastes my time.'' (ID284)
Finally, it was fairly common in the two amotivated profiles ($\sim30\%$) to find the technology 
\textit{meaningless} ($\sim12\%$): e.g., ``\textit{it means nothing except a technology I have to stop using}'' (ID521).  While participants in the \textit{Medium} (18\%) and \textit{External} (14\%) profiles also discussed lack of meaning, their comments tended to be less explicitly negative: often suggesting the technology had pragmatic or escapist value but no deeper personal meaning: e.g., ``\textit{they don't mean anything really, i just like them}.'' (ID147).

\paragraph{Good or Bad for People}
It was common for participants to articulate ways in which the technology was good or bad for them personally, for others, or for society. Users in the \textit{High Quality} profile were most likely to describe ways in which the technology was good for them (in 27\% of accounts): for example, by supporting self care, personal growth, or health and fitness. This was less common in the \textit{Medium} and \textit{Amotivated-Intrinsic} profiles (18\% each) and rarest in the \textit{Amotivated} (7\%) 
 and \textit{External} (5\%) profiles. 

Meanwhile, users in the \textit{Amotivated-Intrinsic} profile were most likely to suggest the technology was bad for people (61\%).
This focused largely on how the technology fed bad habits and compulsion, or diminished mood and self-image. 
Such concerns also featured in the \textit{Amotivated} (35\%) and External (24\%) profiles. In the External profile, however, negative impacts were a little different in character, focusing less on compulsion and more on personal insecurities. For example: \textit{``Despite feeling like the app is harmful to  my mental health, I keep using it because I feel like it might be helpful. I  might land a dream job or connect with someone who can help me excel in my  career. In some ways, it's also to torture myself and compare myself to  others (gives me something to work towards, in an odd way).''}. 

\paragraph{Ascribing Qualities to the Technology}
Participants in the \textit{High Quality} and \textit{Medium} profiles were most likely to ascribe positive qualities to their chosen technologies ($\sim60\%$ vs 30\% or less in other profiles). 
Participants commented on efficiency, convenience and ease of use ($\sim40\%$); the ability to integrate or adapt to their needs ($\sim18\%$, e.g., \textit{``ready to be customised to fit my unique needs. I started small [...] but soon discovered the true power of this app.''} (ID343)), and ways in which the technology was exceptional ($\sim19\%$), whether in outperforming alternatives or continually getting better over time.
Meanwhile, negative qualities were more commonly mentioned in the \textit{External} and \textit{Amotivated} profiles (66\% and 75\%, respectively). 
Both discussed a range of issues including difficulty in use (e.g., ``\textit{For me this programme is really hard to master}'' (ID518)), low or declining quality  (e.g., ``It was a good app at the  begenning but now it is a trash.''  (ID396)), and failure to live up to promises.
Meanwhile, users in the \textit{Amotivated-Intrinsic} profile were least likely to ascribe qualities to the technology, both positive and negative (in 30\% of responses).
Notably, in the two amotivated groups, 
positive qualities 
were often framed as being in conflict with wider problems. One participant, for instance, enjoyed the convenience of face-recognition but felt discomfort at the idea of losing anonymity (ID237). Another was concerned by the effectiveness of a technology which helped their child sleep, fearing the child might become reliant (ID483).
Where users in the High Quality profile addressed technologies they did not feel good about, this often reflected frustrating elements of the technology, in tension with the app's value, e.g., an unreliable, albeit enjoyable streaming app, ``when I am lucky it works on my phone [...], I grab it without looking back.'' (ID 321)

\section{Discussion}

Organismic Integration Theory (OIT) --- a subtheory of SDT --- posits that both the quantity and the quality of motivation play crucial roles in influencing experience, behaviour, and need satisfaction \cite{ryan2017organismic}. 
In this work, we combined OIT and latent profile analysis to better understand the role of motivation in UX. 
Specifically, we identified five \textit{motivational profiles} that captured how users were intrinsically and extrinsically motivated to engage with particular technologies. 
These profiles were associated with significant differences in how users experienced their technology use, and qualitative differences in how users described their engagement. They distinguished identity-grounded and often highly meaningful engagement from empty habit, and empty habit from external pressure.  
While previous work in HCI has neglected extrinsic motivation or treated it as a negative factor \cite{turkay_self-determination_2023, hammer_quality--user-experience_2018,zhu_market_2016, massung_using_2013}, we found that varieties of extrinsic motivation were important in differentiating outcomes, and often associated with positive UX. In particular, autonomous forms of extrinsic motivation were strongly associated with positive affect and need satisfaction. Meanwhile, in our Amotivated-Intrinsic profile, relatively high intrinsic motivation coexisted with negative outcomes: low need satisfaction and compulsive engagement, raising questions around exclusively positive interpretations of intrinsic motivation.
Importantly, the profiles we observed were not technology specific. Some technologies were more prominent in particular profiles, but all profiles contained multiple technology types, and all technology types appeared in multiple profiles (see \autoref{tab:profile_technologies}).
Together, these findings demonstrate how motivational profiling can bring additional nuance to understandings of UX and clarify the conditions which ground and differentiate meaningful and compulsive engagement. 


In the following, we discuss the implications of our findings for theory and practice --- in design, UX, HCI more generally, and also in primary SDT research.

\subsection{Motivational Profiles in User Experience}
\label{skew}
UX research 
has commonly addressed motivation via need satisfaction, investigating how this influences affective experiences, technology evaluations and other positive outcomes  \cite[e.g.,][]{burnell_technology_2023, tuch_leisure_2016, partala_understanding_2015}, and how need satisfaction is itself influenced by factors like use context \cite{tuch_leisure_2016}, and the kind of goal pursued \cite{mekler_momentary_2016}.
In these terms, we find our results are consistent with 
prior findings: participants in the positive condition reported higher mean scores than those in the negative condition on every dimension of need satisfaction, positive affect, usability and self-attribution. This consistency provides an strong baseline for further interpreting results.

Notably, however, we observed larger differences in UX outcomes between motivational profiles than between the positive and negative conditions.
Scores for need satisfaction and affect (PANAS --- often treated as a yardstick of positive UX \cite[e.g.,][]{hassenzahl_needs_2010, tuch_analyzing_2013, hassenzahl_experience-oriented_2015, mekler_momentary_2016}) were often far higher in autonomous profiles than in the positive condition, and far lower in non-autonomous, amotivated, profiles than in the negative condition.
In particular, autonomy need satisfaction scores for adjacent profiles were often separated by more than one scale point --- noteworthy since autonomy satisfaction has often been excluded from analysis in previous UX research \cite[e.g.,][]{hassenzahl_needs_2010, hassenzahl_experience-oriented_2015, mekler_momentary_2016, tuch_leisure_2016}.

Counter to assumptions in previous HCI research \cite[e.g.,][]{turkay_self-determination_2023, zhu_market_2016, massung_using_2013,hammer_quality--user-experience_2018}, need satisfaction and positive affect in our sample  are not more strongly associated with intrinsic motivation than with extrinsic motivation.
While (consistent with SDT propositions \cite[][p.~171]{ryan_self-determination_2017}) we observed that higher
intrinsic motivation was \textit{generally} associated with higher PANAS scores, this was not true in every instance. 
The Amotivated-Intrinsic profile, for example, scores substantially higher on intrinsic motivation than the External profile, yet these profiles differ little in terms of PANAS scores. Elsewhere, differences in outcomes seem weakly related to intrinsic motivation, and more strongly associated with extrinsic motivation. 
The High Quality and Medium profiles, for example, have very similar intrinsic motivation scores, yet display significant and sometimes large differences across all three basic needs, meaning, self-esteem, and PANAS. These profiles differ most substantially in terms of (extrinsic) integrated regulation.

Taken together, these results suggest that researchers interested in undersatnding support factors for positive UX should look beyond the immediate satisfactions intrinsic to the activity, and consider the wider range of factors which help users integrate activities with their values and goals. 

\subsection{Differentiating Aspects of Autonomy Experience}
Recent work has called for HCI researchers to move beyond simple, singular understandings of autonomy and
instead seek to understand  different aspects of human autonomy, and how they 
come together to influence outcomes of technology use \cite{bennett_how_2023, guldenpfennig_autonomy-perspective_2019}.
In this section we discuss how our results contribute to this programme. 


Previous work in HCI and UX has commonly addressed autonomy via autonomy need satisfaction. This concerns episodic and ``spontaneous'' \cite[e.g.,][p.~3932]{deterding_contextual_2016} aspects of autonomy experience, grounded in the user's immediate inclinations and anticipations \cite[][]{ballou_basic_2023, deterding_contextual_2016}.
In contrast to this, autonomy in motivational regulation concerns the degree to which users make an activity their own: how they internalise the values and patterns of behaviour associated with an activity, alongside their personal values, beliefs, and patterns of behaviour \cite{ryan_self-determination_2017}. To give examples, when a technology frames an activity in a way that matches the user's self-image (e.g., as ``a fun person'', ``motivated'', ``reliable'') then it is experienced as autonomous and more likely to be accepted. When the framing challenges the user's self-image, the activity is experienced as non-autonomous and more likely to be rejected \cite[][p.~735]{gerstenberg_designing_2023}.

Our High Quality profile reflects highly autonomous regulation. It has high scores for integrated regulation, indicating strong internalisation. Participants in this group often connected their technologies with personal goals, vocations, and social roles: describing how they appreciated the technology ``as a mom'', ``as a gamer'', etc.. 
However, not all internalisation and autonomous regulation must involve strong connections to personal identity. The High Quality and Medium profiles also had high scores for another form of autonomous regulation: identified regulation. Despite its name, this does not concern identification with the technology or activity itself. Rather it concerns how users identify links between the activity's outcomes, and other things they value. For example, some participants felt their technologies helped them achieve what they wanted, despite having ``no deeper meaning''. Others described intentional and considered use of entertainment technologies for purposive relaxation. 
Integrated and identified regulation thus illustrate two qualitatively different kinds of autonomous regulation; two different ways of relating an activity to one's goals and values. In SDT these are seen as differing in their ``relative autonomy'', a technical term for \textit{autonomous motivation} which concerns their degree of internalisation.
However, on a broader understanding of what it means for a person to be autonomous \cite[see e.g.,][]{killmister_taking_2018, oshana_personal_2016, prunkl_human_2022,colburn_autonomy_2022}, we suggest that neither of these regulations should be seen as inherently more autonomous than the other. Rather, we suggest they reflect \textit{different} ways of being autonomous --- mindful, conscious and self-governing --- in an activity.

Our results also illustrate two qualitatively different kinds of \textit{low} autonomy regulation, where activities \textit{are not} coherently related to the user's values and goals.
First, the External profile reflects low autonomy due to \textit{external control}. This profile has high scores for external regulation -- the regulation associated with the pursuit of rewards (e.g. money, status) and the avoidance of sanctions (e.g. social censure, punishment). Scores for introjected regulation are also high, indicating partial and rigid internalisation of the activity, related to phenomena like guilt and conditional regard. 
This form of low autonomy motivation contrasts with our two \textit{amotivated} profiles. Amotivation is not related to the presence of external control, but rather to the lack of intentionality and internalisation. This is expected to arise either from users' indifference to the activity, or from conditions which undermine either their sense that they are able to perform adequately or 
their expectation that the activity will lead to valued outcomes \cite[][p.~191]{ryan_self-determination_2017}. 
While previous work in UX has not distinguished between these kinds of low-autonomy engagement, they have distinct experiential outcomes in our results. While both are associated with low need satisfaction and PANAS scores, the External profile (reflecting external control) scores far more highly for competence, relatedness and security satisfactions. Participants in the two amotivated profiles, in contrast, were more likely to feel the technology was actively bad for them, and to think of quitting the technology \cite[see also similar results in][]{bruhlmann_measuring_2018}.
In making this comparison it is important to note that the mix of technologies addressed by participants differed across the Amotivated, Amotivated-Intrinsic and External profiles. 
In future it will be important to investigate different kinds of low-autonomy engagement in more technologically-homogeneous samples. In particular, researchers may consider 
social media technologies, which were found in all of our motivational profiles, including those dominated by external control and amotivation. 

\subsection{Intrinsic Motivation and Hedonic Amotivation} \label{understanding intrinsic}

Intrinsic motivation is widely considered an uncomplicated good, grounded in the satisfaction of three basic psychological needs which are essential to human flourishing. Within SDT it is described as ``a prototype of human proactivity that is experienced as highly autonomous'', and an evolved ``growth-oriented potential, which ... characterizes our bases for learning and developing'' \cite[p.~222]{shaver_beyond_2012}. Intrinsic motivation is  considered ``by definition, autonomous; [...] experienced as being volitional and emanating from one’s self'' \cite[p.~14]{ryan_self-determination_2017}.

Results for the low-autonomy Amotivated-Intrinsic profile complicate this picture: The factor score for intrinsic motivation in this profile is quite high: markedly higher than the intrinsic motivation scores in the Amotivated and External profiles (see \autoref{lpa_profiles}), and
equivalent to $\sim5$ out of $7$ on the scale. 
At the same time, the profile has a very high score for amotivation, a construct associated 
with the \textit{absence} of need satisfaction
\cite[][p.~190]{ryan_self-determination_2017}
, and the lack of conscious volition.
These qualities are reflected in the Amotivated-Intrinsic profile's very low need satisfaction scores and in participants' descriptions of compulsion, bad habits, and diminished emotional states in relation to their chosen technologies.
Rather than the `growth-oriented potential'' \cite[p.~222]{shaver_beyond_2012} 
SDT associates with intrinsic motivation, these results paint a picture of immediate but shallow pleasure, and weak self-determination. We call this \textit{hedonic amotivation}\footnote{Following Tamborini et al's distinction between hedonic and nonhedonic satisfactions \cite{tamborini_ron_media_2011}, see \autoref{routesforward} for more detail.}.

In interpreting these results it is important to note that not all psychologists concur with SDT's view of intrinsic motivation. Lindenberg, for example, has argued against SDT's maxim that all intrinsic motivation is grounded in basic need satisfaction, 
\cite{lindenberg_siegwart_intrinsic_2001}. He suggests this framing fails to account for many factors, such as stimulation, which are \emph{intrinsic} to the activity, motivating and enjoyable, yet are unconnected to basic need satisfaction and lack its positive, healthy, associations.
He argues this risks creating gaps in the treatment of unhealthy engagement: ``the very distinction between the inherently noble connotations of intrinsic motivation and obsessive and addictive behaviour has been left unexamined'' \cite[p.~318]{lindenberg_siegwart_intrinsic_2001}. We see this as a particularly salient issue in technology research, where SDT is applied to understand issues like digital self-control \cite[e.g.,][]{lukoff_designing_2022} and problematic patterns of gaming \cite[e.g.,][]{ballou_basic_2023,mills_self-determination_2020}.
Alongside these theoretical concerns, results in a number of SDT-informed studies of technology challenge SDT's unified view of intrinsic motivation (albeit the challenge to SDT is rarely articulated explicitly). 
In a study of online gamers \cite{bruhlmann_motivational_2020}, for example, around a third of participants shared a motivational profile similar to our Amotivated-Intrinsic profile, and likewise reported negative UX outcomes and low need satisfaction.
Elsewhere, Wan and Chiou found that more ``addictive'' behaviour could be associated with higher intrinsic motivation \cite{wan_motivations_2007}, and Mills and Allen found almost no correlation between levels of intrinsic motivation and lower problematic gaming behaviour \cite{mills_exploring_2018}.

While these tensions have not, to our knowledge, been addressed directly in core SDT research, Rigby and Ryan \cite[p.~99]{rigby_glued_2011} articulate a seemingly related phenomenon, subsequently named the ``need density hypothesis'' \cite{ryan2017motivation}. 
The account primarily addresses problematic gaming behaviour, but its principles transfer readily to problematic behaviour with other forms technology, such as social media.
While we see potential issues and limitations in the account (which we articulate below), we suggest it offers a useful starting point when seeking to 
clarify potential gaps in SDT's account of intrinsic motivation
and thereby understand results like those observed in our Amotivated-Intrinsic profile. 
The account is grounded in Rigby and Ryan's observation that, for some players, an activity which gave them  "a lot of happiness \textit{moment to moment} was at the same time creating a dull ache in [their] soul'' \cite[p.~99, our emphasis]{rigby_glued_2011}. While SDT generally expects basic need satisfaction to support positive experience, Rigby and Ryan suggest that in some cases games (and other digital technologies) may have properties which interfere with this. Specifically, they often combine high levels of need satisfaction  
with very low barriers to engagement \cite[pp.~100-105]{rigby_glued_2011} -- 
particularly in comparison to
need-satisfying activities which are not mediated by digital technology. The authors propose that such `need-dense' 
activities may provide a strong draw to people who lack need satisfaction in their wider lives, and may thereby lead to problematic over-reliance \cite{mills_exploring_2018, vansteenkiste_psychological_2013}.

Evidence for the need density hypothesis remains inconclusive at present \cite[][pp.~13-14]{tyack_self_2024}. The results for our Amotivated-Intrinsic profile fit some aspects of the account while raising challenges for other aspects.
Participants in that profile often described the use of entertainment and social media apps: technologies which are generally designed to be enjoyable and present little barrier to engagement. They commonly reported compulsive patterns of behaviour, and had relatively high intrinsic motivation scores.
However, while SDT associates intrinsic motivation with need satisfaction, participants in this profile in fact reported very \textit{low} scores for need satisfaction, alongside high scores for amotivation --- a construct associated with low need satisfaction \cite{bruhlmann_measuring_2018}, and high need \textit{frustration} \cite[][p.~190]{ryan_self-determination_2017}. In short, in this profile compulsive behaviour and low affect are observed alongside relatively dense \textit{intrinsic motivation}, yet \textit{sparse} need satisfaction. This raises questions about the role played by basic need satisfaction in disordered technology use and, more broadly, in intrinsic motivation.


\subsubsection{Routes forward for Understanding Intrinsic Motivation in Technology Use} \label{routesforward}



We see two potential routes forward to clarify the sources and consequences of intrinsic motivation in technology use. We develop these as speculative, though theory-grounded, propositions to direct future research and theory building. 
These proposals not only address technology use but SDT's account of motivation more broadly, insofar as the behaviours and experiences observed during technology use are unlikely to be completely absent in other areas of life.

\paragraph{`Bubbles' of episodic need-satisfaction}

First, we draw attention to the different temporal properties of intrinsic and extrinsic motivation. In SDT's account, internalised extrinsic motivation (see \autoref{OIT_diagram}) relies on satisfactions which do not pay off during the activity itself, but in the outcomes of the activity, and via personal investment over time. As such, extrinsic forms of motivation should be seen as temporally extended and situated in the wider circumstances of a person's life. Ryan and Deci emphasise that truly autonomous extrinsic motivation involves ``bring[ing] the future into the present'' \cite[][p.~198]{ryan_self-determination_2017} --- connecting our immediate activity to the fulfilment of values and goals which are not directly present. By contrast, intrinsic motivation is 
concerned with immediate, \textit{spontaneous}, experience \cite[p.~198]{ryan_self-determination_2017}: ``the freely chosen continuation of an activity in `free time', \textit{measured in seconds}'' \cite[p.~318; our emphasis]{lindenberg_siegwart_intrinsic_2001}. SDT proposes that need satisfaction can support both intrinsic and autonomous extrinsic motivation \cite[][p.~123, 202]{ryan_self-determination_2017}, but we hypothesise that in some circumstances it may primarily support the former, and have little impact on the latter.


In short, we propose that problematic technology use, such as that 
described in the need density hypothesis, may follow from a kind of need satisfaction `bubble': where need satisfaction is high, moment-to-moment, but has little effect beyond the bubble of immediate engagement. In such circumstances competence, autonomy, and relatedness may be satisfied by choices, abilities, and relationships which do not play much of a role outside the interaction, in wider life. This may mean, for example, 
interactions with people who play no role in our life outside social media, or shallow representations of our ethical values which lack meaningful consequence.
Such circumstances may not provide a supportive context for internalising activities --- connecting them to our wider identity, values and goals. 
We suggest this might explain how intrinsic motivation can coincide with amotivation: the latter need not reflect a complete lack of motivation to engage, but can concern a lack of motivational internalisation and intentionality \cite[][p.~190]{ryan_self-determination_2017}.
If this hypothesis is correct, then we would expect to see different results on need satisfaction scores when the activity is considered at different scales of life: Users should report relatively high need satisfaction when considering their immediate experience during the activity, but low need satisfaction when considering the activity in the context of their wider life.

One route forward in testing this ``need bubble'' hypothesis could be to make use of the questionnaires provided by the METUX framework, which address need satisfaction at different scales of behaviour \cite{peters_designing_2018, burnell_technology_2023}. Indeed, we see this hypothesis as somewhat congruent with the account outlined in the METUX framework wherein need satisfaction can differ between "spheres" of life \cite[][see also \cite{ballou_basic_2023} with regards to games and well-being]{peters_designing_2018}.
Our account goes beyond these previous accounts in explicitly articulating a mechanism for this scale-dependency in terms of internalisation, and the temporal properties of motivational constructs.

\paragraph{Sources of intrinsic motivation beyond need satisfaction}
The ``need bubble'' account above seems plausible to us. It would also be consistent with existing SDT propositions, namely that intrinsic motivation follows from basic need satisfaction \cite[][p.~123-178]{ryan_self-determination_2017}, and that need satisfaction can be somewhat restricted to particular spheres of life \cite{peters_designing_2018}. However, we also find another hypothesis compelling: \textit{basic need satisfaction may not be the only source of intrinsic motivation}. Other factors, beyond autonomy, competence and relatedness, may support intrinsic motivation without supporting internalisation. Again this can be expected to promote `bubbles' of motivation which have little positive consequence beyond the immediate episode. 

This hypothesis is at odds with core SDT propositions but, as noted above, it is not without precedent. Previous research in motivational psychology has proposed sub-types within intrinsic motivation, including stimulation \cite{vallerand_academic_1992}, and posited a role for more ``hedonic'' needs, such as arousal \cite{tamborini_ron_media_2011, tamborini_ron_defining_2010} and novelty \cite{gonzalez-cutre_testing_2020}, in addition to the ``basic'' needs for autonomy, competence, and relatedness.
Extending this line of thought, we argue that more hedonic satisfactions such as stimulation, novelty and sensory pleasure may contribute to intrinsic motivation, but not autonomous extrinsic motivation. That is, these may be effective in supporting positive affect and continued engagement, moment-to-moment \cite[see e.g.,][]{kao_how_2024}, but may provide little support for the internalisation of motivation, nor for positive, longer term, outcomes such as increased positive affect and improved well-being. 
We see some initial evidence for this hypothesis in previous UX results: Hassenzahl et al. found that, among Sheldon's 10 needs \cite{sheldon_what_1996}, stimulation was the most strongly predictive of positive affect 
\cite{hassenzahl_needs_2010}. Meanwhile Mekler and Hornbæk found that hedonic (pleasure-seeking) motivations correlated with stimulation, but not basic need satisfaction \cite{mekler_momentary_2016}.

If intrinsic motivation has other sources beyond the three basic needs, then it should be possible to observe relatively high intrinsic motivation in episodes of interaction \textit{even when the interaction does not satisfy the needs for autonomy, competence and relatedness}. 
This contradicts current accounts in SDT, but is consistent with the results observed in our Amotivated-Intrinsic group. Future work may test this hypothesis by measuring the association between intrinsic motivation and constructs like stimulation and novelty. 

Evidence in favour of this hypothesis would have important consequences for the interpretation of intrinsic motivation measures in HCI, UX, and primary SDT research. Generally, self-report of intrinsic motivation is operationalised via adjectives like "enjoyable", "interesting", and "fun" \cite[e.g.,][]{buil_encouraging_2019, bruhlmann_measuring_2018}, which seem unlikely to distinguish between basic need satisfaction and hedonic aspects of the experience such as stimulation and novelty.
It is not clear that these hedonic aspects of experience should be expected to  support the kind of ``growth-oriented potential, [...] learning and developing'' \cite[p.~222]{shaver_beyond_2012} associated with basic need satisfaction. As such, we suggest researchers should be cautious about connecting intrinsic motivation to positive consequences for well-being and sustainable motivation.

As an interim solution we suggest that researchers measuring intrinsic motivation might also consider measuring extrinsic motivations. As our results illustrate, extrinsic motivation and amotivation provide a useful context for interpreting the intrinsic motivation score. The degree of congruence between the intrinsic and extrinsic constructs can help clarify when intrinsic motivation scores are or are not grounded in basic need satisfaction. 

Finally, we note that the separation of basic needs from hedonic satisfactions seems likely to be more common in technology use than in other contexts. Digital technologies allow stimulation to be precisely calibrated, highly responsive to the user, and somewhat detached from wider consequences, and entertainment design in particular is known to leverage such possibilities \cite{kao_how_2024}. 
As such this is an aspect of human motivation which HCI research may be uniquely placed to address. The discipline is well positioned to ``talk back'' \cite{beck_examining_2016,tyack_self_2024} to SDT, by understanding motivational phenomena which also have ramifications for core aspects of SDT theory.

\subsection{Design for Internalisation}\label{sec:design_internalisation}
Gerstenberg and colleagues recently argued that OIT provides ``an ecological perspective that integrates [...] activity into the broader context of a person's life'' \cite[][p.~727]{gerstenberg_designing_2023}, which designers can draw upon to support healthy sustainable motivation \cite{gerstenberg_designing_2023}.  
They developed an OIT-grounded approach to design for behaviour change which they call "technology-mediated regulatory transformation". 
Their work focuses in particular on guiding users to more autonomous forms of regulation in rehabilitation exercises, to support more sustainable engagement \cite{gerstenberg_designing_2023}.
Our results support the idea of grounding design in OIT, as well as supporting certain details of the approach Gerstenberg et al. describe. At the same time we see more expansive opportunities going beyond behaviour change technologies, to address UX design more generally. We also see certain situations where propositions in Gerstenberg and colleagues framework need to be clarified or tested in order to support future evidence-based design work.
In this section, we propose a research agenda for motivational design, grounded in OIT and our results. We call this \textit{Design for Internalisation}, since it aims 
to support users in moving toward more internalised, autonomous forms of motivation.

 \begin{figure}[t]
  \includegraphics[width=1\linewidth]{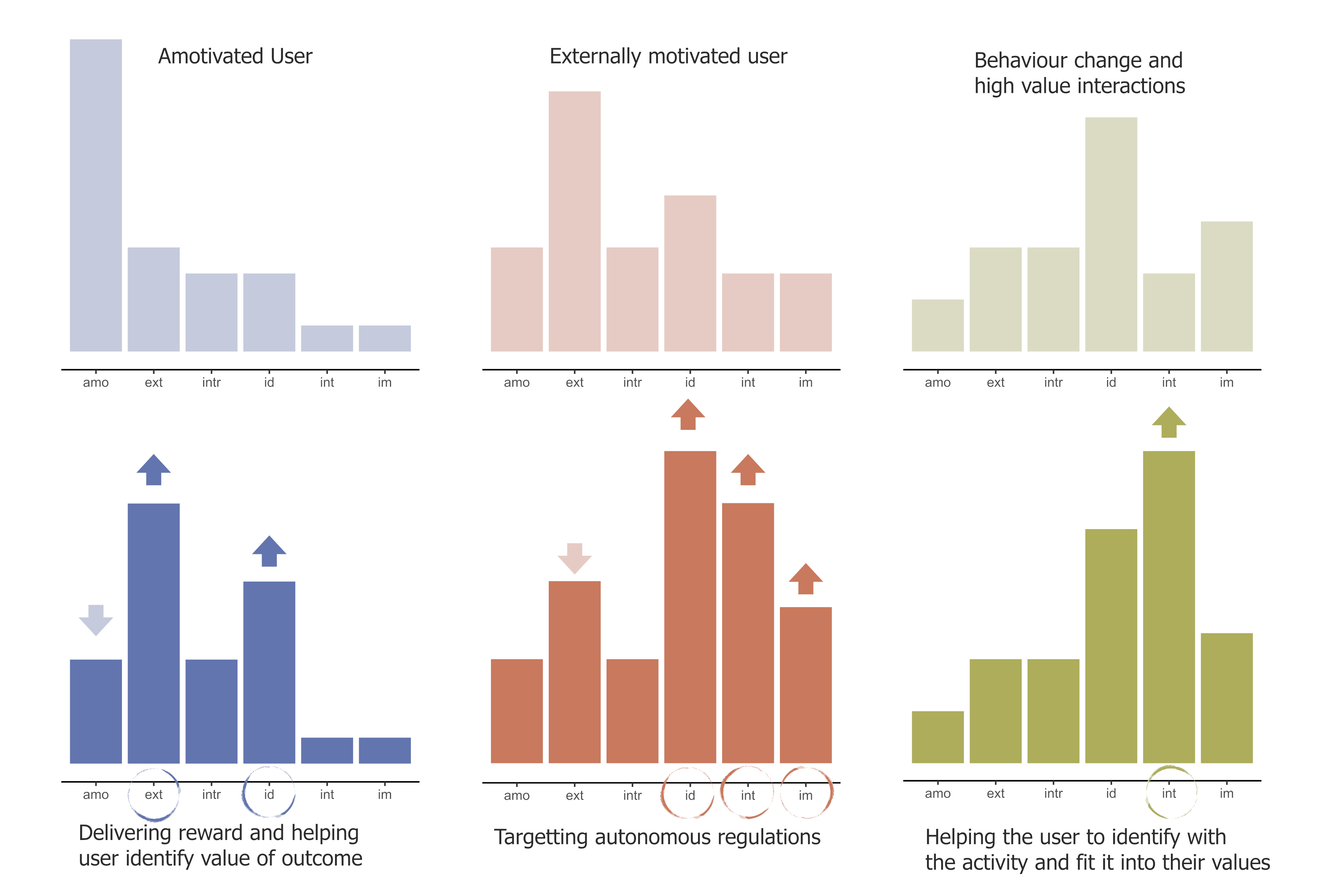}  
  \caption{Three possible approaches to design for motivational internalisation, calibrated to users' existing motivational profiles. \textbf{Left} --- For amotivated users improved experience and engagement may be supported by careful use of reward, and by clarifying the basic value of the activity's outcomes for the user. This may provide a baseline of motivation for further internalisation. \textbf{Middle} --- For users with less amotivation, but whose motivation is reward-focused, it may be possible to support healthier, more sustainable engagement. Designers may address the value of the activity's outcomes relative to the user's goals, its intrinsic pleasures, or address how the activity relates to the user's identity, social roles, etc. \textbf{Right}: for activities of high personal significance, designers might provide the conditions that support self reflection and identification with the activity, and the integration of the activity with goals and values --- what Gerstenberg and colleagues call ``design for \textit{integration}'' \cite{gerstenberg_designing_2023} }
  \label{fig_internalisation}
 \end{figure}

\subsubsection{Evidence for internalisation strategies}
First, our empirical results provide an evidential basis for design strategies to support more autonomous forms of motivation --- such as those proposed by Gerstenberg and colleagues \cite{gerstenberg_designing_2023}. 
Our High Quality profile, for example, is dominated by integrated regulation --- the goal of Gerstenberg et al.'s design strategies. 
Compared to other profiles, users in this profile reported greater meaning, need satisfaction and positive affect, and were more likely to connect the technology to their identity and values. This is consistent with Gerstenberg et al.'s strategy of helping users see the fit between the activity and their identity (e.g., their sense of having a good memory, being fun, etc.). 
Our results also suggest other strategies to support integration: Participants in  
the High Quality profile emphasised creative aspects of their technologies and 
connected them to their social roles (e.g. ``as a mom'', ``...gamer'', ``...tech guy''). Designers might therefore consider relating activities and technologies to users' social roles and identities, or introducing self-expressive and creative activities.  
More broadly, since both reflection and awareness play an important role in integrated regulation, it seems likely that there is value in design strategies which promote these mindsets, such as seamfulness \cite{inman_beautiful_2019}, friction in interaction \cite{cox_design_2016, mejtoft_design_2019} and varieties of positive breakdown \cite{bennett_multifractal_2022, bennett_jank_2023}.

Our results also support the idea that designers should not limit themselves to integrated regulation, but address the full spectrum of extrinsic and intrinsic motivation.
While integrated regulation is considered the most autonomous and well-being supportive form of extrinsic motivation \cite{ryan2017organismic},  it is inherently difficult to support this through design. Integration relies on developing congruence between the activity and the user's self image, and as such can require significant investment from the user; sometimes even emotional investment in the form of ``a confrontation with personally held values'' \cite[][p.~5]{gerstenberg_designing_2023}. 
While it may be reasonable to expect such high levels of investment from the user in matters of health and high personal significance, this cannot be relied upon in other circumstances. 
Further, the self-confrontations associated with integration may be risky, requiring contextual, interpersonal or therapeutic support, to avoid backfiring and undermining motivation or self-image. Designers can rarely ensure such support.

In general, our results suggest that designers should largely focus on more intermediate changes in users' motivational profiles. 
As illustrated in \autoref{fig_internalisation}, our results suggest that improved need satisfaction may often be supported by \textit{any} internalisation of motivation: that is, by any rightward shift in the motivational profile towards more autonomous regulation. This may, for example, mean a shift away from amotivation, by increasing either intrinsic or extrinsic motivation. Equally it may mean a shift away from externally controlled motivation --- based on reward, social pressure, or guilt --- and toward more autonomous regulation. 
Such intermediate strategies for internalisation may entail clarifying the value of an activity's outcomes, and connecting these outcome to the user's goals (identified regulation as in \autoref{fig_internalisation}, middle). Gerstenberg et al. provide some discussion of such design strategies \cite{gerstenberg_designing_2024}. Beyond these, we suggest that in certain cases calibrated reward may also be effective (external regulation as in \autoref{fig_internalisation}, left). 

In developing such strategies for motivational design, our results (and OIT more broadly) suggest that users' responses to such interventions may be affected by the context, by their existing motivational profile for the activity, and by nuances in design \cite[e.g., in the kind of reward and the way it is delivered,][p.~148]{ryan_self-determination_2017}. Since these issues may be important in supporting healthy and positive outcomes, we articulate them further below.

\subsubsection{How do motivational profiles change?}

When designing for internalisation of motivation it will be important to understand how users move between profiles over time, and to understand the factors that support such change. 
OIT is vague about what happens when motivational regulation changes \cite{gerstenberg_designing_2023}, and there has been a surprising lack of research on the dynamics of motivation. 

Gerstenberg et al.'s work \cite{gerstenberg_designing_2024}, discussed above, is among the few works to investigate the factors which facilitate internalisation. 
However, their approach is grounded in some assumptions which are in tension with OIT propositions, and which currently lack an evidential basis.  Specifically, the approach assumes: 1) A user's motivation for an activity will be dominated by one motivational regulation. 2) Users will move step-wise, between adjacent motivational regulations. 3) These changes in regulation will provoke an experience of ``friction'' between the two adjacent regulations, which the user must overcome. 
Counter to these assumptions, OIT holds that multiple motivational regulations will commonly be operative for a single activity, simultaneously \cite[][p.~184]{ryan_self-determination_2017}. Our own results suggest that autonomous regulations can happily coexist, and longitudinal studies have shown that less autonomous regulations need not decline as more autonomous forms of motivation increase \cite[][]{wasserkampf_organismic_2016}.  
Below we outline open questions and theoretical propositions which can guide future work to understand design for internalisation.

First, we suggest that \textbf{motivational strategies should be sensitive to the user's current motivational profile} for the activity. Internalisation will often demand engagement and effort from the user, and as such seems likely to demand some existing level of motivation. We should not, for example, expect an amotivated user to engage in the ``confrontation with personally held values'' \cite[][p.~5]{gerstenberg_designing_2023} required for integration. For users high in amotivation it may be more effective to focus on clarifying the value of the activity's outcomes (identified regulation), their relevance to the users' goals and values, or to seek ways to make the activity more immediately enjoyable (intrinsic motivation).
While previous work has often treated external reward as straightforwardly negative \cite[e.g.,][]{zhu_market_2016, massung_using_2013, hammer_position_2017},  we suggest that this may depend on the user's existing motivational profile. There may be value in the use of rewards for users whose profiles are dominated by amotivation or external regulation \cite[][pp.~148]{ryan_self-determination_2017}\footnote{It is important to note that SDT indicates the kind of reward and the manner of its delivery may be critical. ``Informational'' and ``tangible'' rewards are expected to have quite different motivational consequences \cite[][p.~148]{ryan_self-determination_2017} though it is not always clear how these categories map to concrete reward types used in digital environments.}, since this may support initial engagement which later allows for internalisation. Meanwhile, for users with different motivational profiles, rewards may be counterproductive.
For example, users who identify strongly with an activity (e.g., high integration) may perceive the introduction of reward as `cheapening' the experience, undermining the congruence of the activity with their values. 


Related to this first point, in some cases there may be value in \textbf{addressing motivation as a ``ladder''}. The motivational regulations in OIT are not developmental: people are not expected to move through them stage by stage \cite[][p.~199]{ryan_self-determination_2017}, and in principle should be able to directly develop any of the regulations. 
However, if the process of internalisation itself makes demands on motivation then this suggests that a stepped or incremental approach to motivational design may be effective. 
There is thus some sense in treating new adopters and low motivation users as beginning at the bottom of a motivational ``ladder'', with little existing 
motivation for designers to draw upon. Here designers may be best served by strategies which are likely to work quickly and require little investment from the user --- for example, reward, novelty, and immediately enjoyable interactions.
The motivation generated by such approaches might, in turn, 
be leveraged to support motivational strategies which require more investment from the user. This might include prompts to reflect on the value of outcomes, or questionnaires which help the system to align with the users' values and goals. In this way, designers might seek to move the user toward progressively more autonomous and sustainable forms of motivation. 
Related ideas about the staged adoption of behaviour can be found in previous work in sports psychology: Wasserkampf and Kleinert argue that while less autonomous motivations such as introjection may be unhealthy and unsustainable in the long term, they may still have a functional role during the initial stages of adoption \cite{wasserkampf_organismic_2016}. 
This account is somewhat in line with Gerstenberg et al.'s own developmental approach \cite{gerstenberg_designing_2024}, but differs in its basic assumptions in ways which have consequences for design. We do not assume that a user's behaviour will be dominated by one regulation, nor that change in profile will be restricted to step-wise change between adjacent regulations. Finally, we do not assume that these transitions will provoke experiential ``friction'' which the user must overcome. We only assume that existing motivation is a pre-requisite for further internalisation. We suggest this is more parsimonious and opens a wider space of design strategies. However, future work may test the different assumptions of these two approaches.

HCI researchers have argued that controlled motivations, such as reward, can undermine integration and intrinsic motivation, leading to poor quality motivation \cite[e.g.,][]{gerstenberg_designing_2023,turkay_self-determination_2023, hammer_quality--user-experience_2018}. This raises the question whether these kinds of motivation might inhibit users from continuing to internalise behaviours and activities. For example: might some rewards encourage the user to rely on those rewards for motivation? If so, might this discourage the user from more reflective engagement with the value of the activity and how it fits into their life: activities which may lead to more autonomous motivation? 
At present there is very little data to answer this. However, in some cases less autonomous regulations \textit{do not} seem to prevent the development of more autonomous motivations: there is evidence that controlled and autonomous regulations coexist, and that controlled regulations remain in place as autonomous regulations develop \cite{wasserkampf_organismic_2016}.
Our own results point to similar conclusions: even in the most integrated, High Quality, profile we observed relatively high introjection.
One possibility is that it is not so much the presence as the \textit{absence} of certain motivations may form a bottleneck to further development. For example, users with very low identified regulation will fail to see how an activity leads to outcomes they value, something which may be difficult to overcome even if the activity itself is quite fun \footnote{We thank our reviewers for this suggestion.}.
Tentatively, we can say that controlled motivations may not \textit{always} inhibit the development of autonomous motivation. Future work should clarify this and identify the contexts in which reward and other controlling motivations do (and do not) hamper internalisation and the development of sustainable, autonomous, motivation. 

Finally, we suggest that designers and researchers \textbf{attend to the timescales over which motivational regulations change}. 
Motivational profiles are unlikely to change on the short timescales characteristic of many HCI studies. While recent work has observed that internalisation of motivation can sometimes occur quite quickly, ``quickly'' here means weeks rather than hours \cite{wasserkampf_organismic_2016}. 
Further, 
the 6 regulations may change over different timescales. Wasserkampf and Kleinert found that some of the extrinsic regulations could increase quickly, while change in intrinsic motivation was slower --- i.e., it took longer before participants began to find the activity inherently interesting and enjoyable \cite{wasserkampf_organismic_2016}. The rate at which regulations change seems likely to be activity-dependent; Wasserkampf and Kleinert's work pertains to exercise motivation, where increase in physical fitness may be required to find the activity enjoyable. 
We recommend researchers consider the impact of timescale both when planning studies and evaluations, and when interpreting results. Studies of short duration might otherwise only observe changes in the faster-changing regulations, and miss slower but still significant changes, something which risks biasing future findings. 

\paragraph{Identifying Profiles}
At their simplest, the strategies discussed above point to some broad rules of thumb --- for example, that designers might use different motivational strategies for new versus established users. However, to make motivational interventions adapt effectively, and to understand the impact of interventions, it will be valuable 
to \textbf{identify individual users' actual motivational profiles}. These approaches may build on the methodology in this paper, using the UMI scale \cite{bruhlmann_measuring_2018} and latent profile analysis to identify particular user groups around which interventions can be tailored. Studies might begin by using LPA to identify subgroups with different motivational profiles and then testing interventions on each subgroup.
Such approaches may also bring nuance into existing user testing processes in industry UX teams, particularly for large user bases.
In future the application of the design principles described above might be supported by the identification of behavioural markers for motivational type. This would allow broad motivational types to be inferred without the need to administer self-report questionnaires. Our analysis of participants' descriptions offers some clues as to candidates for these behavioural markers, e.g., ``lurking'' behaviour on social media (in our study this was associated with amotivation), and a focus of interest and active searching in media consumption (more autonomous motivation).

 \section{Limitations and Future Work} \label{sec:limit_future}


In this study, we tested the high-level hypothesis that differences in motivational profile would be associated with significant differences in UX and need satisfaction.
We then conducted exploratory analyses to understand factors and outcomes associated with particular motivational profiles.
We did not form or test hypotheses about the specific motivational profile that would be observed, nor about relationships between specific profiles and specific outcomes. These aspects of our analysis were exploratory, and in reporting them we have been careful to avoid final or confirmatory claims. 
Below we discuss limitations in our approach and further opportunities our work opens for future work.

First, the profiles we identify should not be treated as a definitive typology of motivation in technology use.
In our study participants were given free choice in selecting the technologies to address.
We considered this a necessary initial step in applying a new approach to UX, making our results comparable with previous critical incident studies \cite[e.g.,][]{hassenzahl_needs_2010, hassenzahl_experience-oriented_2015, tuch_leisure_2016} and providing a broad agenda for future work.  
However, this also resulted in some technologies being addressed by relatively few participants, potentially limiting our ability to resolve representative distributions of motivations for every technology type.
Indeed, when evaluating LPA models we saw some evidence that a larger sample size may have supported a larger number of profiles.  

Future work may therefore focus on particular technology types and contexts. 
This would help identify whether certain motivational profiles are particularly common in certain technology types, allow researchers to investigate contextual factors which shape motivation in particular kinds of technology use, and allow the impact of technology type to be isolated from the impact of motivational profile.
Social media use is a strong candidate for future investigation: these technologies were strongly represented in each of our profiles, indicating that motivational profiles capture relevant differences in types of social media engagement. Online gaming was less well represented in our sample yet, as discussed in \autoref{understanding intrinsic}, there is value in clarifying the role played by different motivational factors in cases where players persist in disempowering and often negative play experiences \cite[see e.g.,][]{vornhagen_im_2023, bruhlmann_motivational_2020}.
Other work might focus on more utilitarian technologies such as workplace technologies and security applications, where intrinsic motivation and need satisfaction might be expected to be low and extrinsic motivations to have a stronger influence. 

It is worth noting that our results may not include very negative experiences with technologies.
In our negative condition participants described technologies they did ``not feel good'' about, which is not necessarily the same as feeling bad. 
We framed the question in this way for two reasons: 1) In order to avoid a polarised sample, capturing extremes of experience. 2) Because we felt that regular and continuous use of technologies that people felt actively bad about would occur under relatively unusual conditions, that might not be representative of the majority of technology use.
Nonetheless,
it remains for future work to understand the motivational profiles which characterise the most negative experiences with technology.

Another issue we could not address in great depth was influence of interpersonal factors on motivational profiles.
While interpersonal factors are expected to be important in motivation and internalisation \cite[pp.~158-216]{ryan_self-determination_2017}, the results of our survey did not support in-depth analysis of this. Nonetheless, our results do indicate interesting directions forward. A number of participants reported that they were introduced to technologies by friends, or adopted technologies together with friends. It is also interesting that the mostly controlled, \textit{External}, profile nonetheless showed quite high scores for relatedness and popularity, despite low need satisfaction in other respects.
Future work might investigate how interpersonal relationships influence motivational profiles. This may focus, for example, on different forms of interpersonal interaction such as balanced two-way communication, unbalanced ``following'' of others' content, and participation in community.
Similarly, our dataset did not permit the analysis of differences in motivational profile over age groups. SDT holds that early childhood motivation is dominated by intrinsic motivation \cite[][p.~214]{ryan_self-determination_2017} but that even by late childhood people are able to access all regulations \cite[][p.~197]{ryan_self-determination_2017}. 
Future work may therefore consider how motivational profiles are associated with stages of life. 

Finally, it is important to note that our results do not provide evidence of causal relationships between motivational profiles, basic needs and other outcome measures. SDT suggests that motivational processes, needs, and identity factors may be related by reciprocal relationships rather than by single clear directions of causality. For example, internalised, autonomous regulations are expected to support need satisfaction and other positive affect \cite[pp.~202-207]{ryan_self-determination_2017}, while at the same time need satisfaction and affect can act as support factors for internalisation \cite[pp.~208-212]{ryan_self-determination_2017}. 
Sheldon and Houser-Marko point to the potential for an ``upward spiral'' of motivation \cite{sheldon_self-concordance_2001}, yet there is evidence that such reciprocal relationships between constructs can also result in more complex, and not necessarily positive, trajectories \cite{heino_studying_2021}. 
Recent work has argued that many experiences in HCI may show complexity of this kind \cite{bennett_emergent_2021, sarsenbayeva_does_2020, berkel_modeling_2020}, and has turned to complexity methods to disentangle the directionality and relative strength of causal relationships between experiential constructs \cite{heino_studying_2021, sarsenbayeva_does_2020, berkel_modeling_2020}. We see great value in future research applying such analyses to understand the network of relationships between the motivational and experiential constructs addressed in this paper.



 \section{Conclusion} \label{sec:conclusion}

Motivation is a key aspect of human experience and behaviour, central to accounts of human psychological thriving. Despite this, previous work in HCI and UX has addressed motivation in a limited fashion, largely focusing on the immediate satisfaction inherent in moment to moment engagement with technologies. Very little attention has been paid to motivational factors related to outcomes of the activity, and how these relate to the values and goals the users bring to the interaction. Yet psychological research suggests that some ``extrinsic'' motivations can be highly supportive of autonomous engagement, and can connect activities to experiences of personal meaning and value. 

We combined Organismic Integration Theory with latent profile analysis to identify five distinct profiles of motivation across a wide range of technology types. These motivational profiles --- capturing intrinsic and extrinsic motivation and amotivation --- were associated with significant and often large differences across key UX outcomes, including need satisfaction, affect, and usability. Our participants' own descriptions of their experiences were then used to shed further light on wider contextual factors associated with these differences in motivation and UX.

Some of our results challenge previous expectations in HCI and SDT. 
Where previous HCI research has ignored or devalued extrinsic motivation, our results indicate that extrinsic motivation can be strongly associated with positive outcomes. 
Where previous research has treated intrinsic motivation as an uncomplicated good, our results indicate that it may in some circumstances be associated with negative outcomes, even potentially compulsive behaviour. 
We relate these results to SDT propositions, articulating new hypotheses and open questions for future work. In particular, we articulate the concept of ``hedonic amotivation'': low-intentionality, compulsive, technology use, which is nevertheless marked by intrinsic motivation. 

Beyond our results, our methodology has great promise as a foundation for a more person-centred approach to quantitative UX research and design. Our approach allows the identification of user sub-groups marked by particular patterns of motivation, who 
should show qualitatively distinct patterns of experience, and who may be receptive to different 
design interventions.  Finally, we articulate a research agenda for 
``design for internalisation'', a person-centred approach to motivational design, aimed at growing positive, healthy, and sustainable engagement.

\begin{acks}
Funded by the European Union (ERC, THEORYCRAFT, 101043198). Views and opinions expressed are however those of the authors only and do not necessarily reflect those of the European Union or the European Research Council Executive Agency. Neither the European Union nor the granting authority can be held responsible for them. 
\end{acks}

\bibliographystyle{ACM-Reference-Format}
\bibliography{references, additional_references.bib}

\appendix

\newpage
\section{Appendix A} \label{appA}

\begin{table}[ht]
\caption{Reasons for excluding participants before analysis}\label{exclusionReasons}
\vspace{-1em}
\begin{tabular}{ll}
Invalid response on one or more check question        & 58 \\
Statistical evidence of careless responding      & 5  \\
Response did not comply with instructions      & 5  \\
Reported poor standard of English language & 2  \\
Reported insincere engagement              & 2  \\
Reported LLM use                           & 3  \\
Very low engagement on open questions      & 4 
\end{tabular}
\end{table}

 \begin{table}[ht]
\caption{Descriptive Statistics for UMI scores at dataset and condition level.  M = mean, CI = confidence interval (lower, upper), SD = standard deviation.}
\vspace{-1em}
\label{tab:umi_descriptive}
\centering
\begin{tabular}[t]{lccccccccc}
\toprule
\multicolumn{1}{c}{ } & \multicolumn{3}{c}{All} & \multicolumn{3}{c}{Positive} & \multicolumn{3}{c}{Negative} \\
\cmidrule(l{3pt}r{3pt}){2-4} \cmidrule(l{3pt}r{3pt}){5-7} \cmidrule(l{3pt}r{3pt}){8-10}
  & M & CI & SD & M & CI & SD & M & CI & SD\\
\midrule
\cellcolor{gray!6}{amotivated} & \cellcolor{gray!6}{2.94} & \cellcolor{gray!6}{[2.78, 3.1]} & \cellcolor{gray!6}{1.85} & \cellcolor{gray!6}{1.73} & \cellcolor{gray!6}{[1.64, 1.82]} & \cellcolor{gray!6}{1.01} & \cellcolor{gray!6}{4.21} & \cellcolor{gray!6}{[4.06, 4.36]} & \cellcolor{gray!6}{1.67}\\
external & 2.17 & [2.04, 2.3] & 1.53 & 1.76 & [1.66, 1.86] & 1.17 & 2.60 & [2.45, 2.75] & 1.73\\
\cellcolor{gray!6}{introjected} & \cellcolor{gray!6}{2.47} & \cellcolor{gray!6}{[2.34, 2.6]} & \cellcolor{gray!6}{1.52} & \cellcolor{gray!6}{2.48} & \cellcolor{gray!6}{[2.35, 2.61]} & \cellcolor{gray!6}{1.50} & \cellcolor{gray!6}{2.47} & \cellcolor{gray!6}{[2.33, 2.61]} & \cellcolor{gray!6}{1.55}\\
identified & 4.73 & [4.59, 4.87] & 1.56 & 5.34 & [5.23, 5.45] & 1.25 & 4.09 & [3.95, 4.23] & 1.60\\
\cellcolor{gray!6}{integrated} & \cellcolor{gray!6}{3.20} & \cellcolor{gray!6}{[3.06, 3.34]} & \cellcolor{gray!6}{1.61} & \cellcolor{gray!6}{3.80} & \cellcolor{gray!6}{[3.67, 3.93]} & \cellcolor{gray!6}{1.50} & \cellcolor{gray!6}{2.56} & \cellcolor{gray!6}{[2.43, 2.69]} & \cellcolor{gray!6}{1.49}\\
intrinsic & 4.92 & [4.77, 5.07] & 1.70 & 5.81 & [5.72, 5.9] & 1.03 & 3.99 & [3.83, 4.15] & 1.76\\
\bottomrule
\end{tabular}
\end{table}

\begin{table}[ht]
\caption{\label{tab:tech_list}Technologies chosen by participants}
\centering
\begin{tabular}[t]{ll}
\toprule
Technology Name & Count\\
\midrule
\cellcolor{gray!6}{instagram} & \cellcolor{gray!6}{33}\\
tiktok & 27\\
\cellcolor{gray!6}{facebook} & \cellcolor{gray!6}{20}\\
youtube & 19\\
\cellcolor{gray!6}{whatsapp} & \cellcolor{gray!6}{16}\\
twitter & 15\\
\cellcolor{gray!6}{reddit} & \cellcolor{gray!6}{10}\\
chatgpt & 8\\
\cellcolor{gray!6}{google maps} & \cellcolor{gray!6}{7}\\
tik tok & 7\\
\cellcolor{gray!6}{discord} & \cellcolor{gray!6}{6}\\
spotify & 6\\
\cellcolor{gray!6}{banking app} & \cellcolor{gray!6}{5}\\
computer & 4\\
\cellcolor{gray!6}{instagram app} & \cellcolor{gray!6}{4}\\
siri & 4\\
\cellcolor{gray!6}{alexa} & \cellcolor{gray!6}{3}\\
amazon & 3\\
\cellcolor{gray!6}{chat gpt} & \cellcolor{gray!6}{3}\\
excel & 3\\
\bottomrule
\end{tabular}
\quad
\centering
\begin{tabular}[t]{ll}
\toprule
Technology Category & Count\\
\midrule
\cellcolor{gray!6}{social media} & \cellcolor{gray!6}{154}\\
streaming media & 44\\
\cellcolor{gray!6}{Messaging and Email} & \cellcolor{gray!6}{34}\\
creativity, media, writing & 32\\
\cellcolor{gray!6}{Computers, phones, tablets} & \cellcolor{gray!6}{21}\\
voice assistants & 21\\
\cellcolor{gray!6}{finance} & \cellcolor{gray!6}{20}\\
workplace technologies & 20\\
\cellcolor{gray!6}{gaming} & \cellcolor{gray!6}{19}\\
Fitness, health and wellbeing & 16\\
\cellcolor{gray!6}{AI Chatbots} & \cellcolor{gray!6}{14}\\
information and maps & 13\\
\cellcolor{gray!6}{other} & \cellcolor{gray!6}{12}\\
shopping & 12\\
\cellcolor{gray!6}{learning and education} & \cellcolor{gray!6}{11}\\
other entertainment & 11\\
\cellcolor{gray!6}{other hardware} & \cellcolor{gray!6}{7}\\
software development and analysis & 6\\
\cellcolor{gray!6}{browsers} & \cellcolor{gray!6}{5}\\
operating system & 5\\
\bottomrule
\end{tabular}
\end{table}

\begin{landscape}
\begin{table}
\caption{\label{tab:bns_var_appendix}Basic Needs Satisfaction variables per profile. M = mean, CI = confidence interval (lower, upper), SD = standard deviation.}
\centering
\resizebox{\linewidth}{!}{
\begin{tabular}[t]{>{\raggedright\arraybackslash}p{9em}rrcccccccccccc}
\toprule
\multicolumn{3}{c}{ } & \multicolumn{3}{c}{Autonomy} & \multicolumn{3}{c}{Competence} & \multicolumn{3}{c}{Relatedness} & \multicolumn{3}{c}{Total Basic Needs} \\
\cmidrule(l{3pt}r{3pt}){4-6} \cmidrule(l{3pt}r{3pt}){7-9} \cmidrule(l{3pt}r{3pt}){10-12} \cmidrule(l{3pt}r{3pt}){13-15}
 & Participants & \% Positive & M & CI & SD & M & CI & SD & M & CI & SD & M & CI & SD\\
\midrule
\textbf{\cellcolor{gray!6}{Mean Positive}} & \cellcolor{gray!6}{254 (51\%)} & \cellcolor{gray!6}{100.0} & \cellcolor{gray!6}{4.75} & \cellcolor{gray!6}{{}[4.67, 4.83]} & \cellcolor{gray!6}{1.32} & \cellcolor{gray!6}{4.31} & \cellcolor{gray!6}{{}[4.21, 4.42]} & \cellcolor{gray!6}{1.73} & \cellcolor{gray!6}{3.61} & \cellcolor{gray!6}{{}[3.5, 3.72]} & \cellcolor{gray!6}{1.85} & \cellcolor{gray!6}{4.23} & \cellcolor{gray!6}{{}[4.15, 4.3]} & \cellcolor{gray!6}{1.20}\\
\textbf{Mean Negative} & 243 (49\%) & 0.0 & 3.38 & {}[3.29, 3.48] & 1.53 & 3.25 & {}[3.15, 3.35] & 1.64 & 2.93 & {}[2.82, 3.03] & 1.74 & 3.19 & {}[3.11, 3.26] & 1.27\\
\midrule
\textbf{\cellcolor{gray!6}{High Quality}} & \cellcolor{gray!6}{165 (33.2\%)} & \cellcolor{gray!6}{74.5} & \cellcolor{gray!6}{5.47} & \cellcolor{gray!6}{{}[5.34, 5.6]} & \cellcolor{gray!6}{0.87} & \cellcolor{gray!6}{5.07} & \cellcolor{gray!6}{{}[4.86, 5.28]} & \cellcolor{gray!6}{1.36} & \cellcolor{gray!6}{4.09} & \cellcolor{gray!6}{{}[3.82, 4.35]} & \cellcolor{gray!6}{1.74} & \cellcolor{gray!6}{4.88} & \cellcolor{gray!6}{{}[4.81, 4.94]} & \cellcolor{gray!6}{0.89}\\
\textbf{Medium} & 146 (29.4\%) & 76.0 & 3.98 & {}[3.79, 4.18] & 1.21 & 3.54 & {}[3.28, 3.81] & 1.65 & 2.97 & {}[2.69, 3.25] & 1.74 & 3.50 & {}[3.42, 3.57] & 0.95\\
\textbf{\cellcolor{gray!6}{External}} & \cellcolor{gray!6}{58 (11.7\%)} & \cellcolor{gray!6}{19.0} & \cellcolor{gray!6}{3.26} & \cellcolor{gray!6}{{}[2.85, 3.66]} & \cellcolor{gray!6}{1.57} & \cellcolor{gray!6}{3.86} & \cellcolor{gray!6}{{}[3.47, 4.25]} & \cellcolor{gray!6}{1.50} & \cellcolor{gray!6}{3.80} & \cellcolor{gray!6}{{}[3.28, 4.32]} & \cellcolor{gray!6}{1.99} & \cellcolor{gray!6}{3.64} & \cellcolor{gray!6}{{}[3.47, 3.8]} & \cellcolor{gray!6}{1.30}\\
\textbf{Amotivated I.M.} & 71 (14.3\%) & 9.9 & 3.07 & {}[2.78, 3.35] & 1.23 & 2.08 & {}[1.86, 2.3] & 0.94 & 2.72 & {}[2.39, 3.05] & 1.42 & 2.62 & {}[2.52, 2.72] & 0.88\\
\textbf{\cellcolor{gray!6}{Amotivated}} & \cellcolor{gray!6}{57 (11.5\%)} & \cellcolor{gray!6}{3.5} & \cellcolor{gray!6}{2.34} & \cellcolor{gray!6}{{}[2.09, 2.59]} & \cellcolor{gray!6}{0.95} & \cellcolor{gray!6}{2.79} & \cellcolor{gray!6}{{}[2.43, 3.15]} & \cellcolor{gray!6}{1.36} & \cellcolor{gray!6}{1.84} & \cellcolor{gray!6}{{}[1.54, 2.14]} & \cellcolor{gray!6}{1.13} & \cellcolor{gray!6}{2.32} & \cellcolor{gray!6}{{}[2.2, 2.45]} & \cellcolor{gray!6}{0.97}\\
\bottomrule
\end{tabular}}
\end{table}

\begin{table}
\caption{\label{tab:additional_needs_var_appendix}Additional Needs Satisfaction variables per profile. "Meaning" shortens "Self-actualisation-meaning, "Popularity" shortens "Popularity-influence" }
\centering
\resizebox{\linewidth}{!}{
\begin{tabular}[t]{>{\raggedright\arraybackslash}p{9em}rrcccccccccccccccccc}
\toprule
\multicolumn{3}{c}{ } & \multicolumn{3}{c}{Meaning} & \multicolumn{3}{c}{Pleasure} & \multicolumn{3}{c}{Security} & \multicolumn{3}{c}{Self-esteem} & \multicolumn{3}{c}{Popularity} & \multicolumn{3}{c}{Total All Needs} \\
\cmidrule(l{3pt}r{3pt}){4-6} \cmidrule(l{3pt}r{3pt}){7-9} \cmidrule(l{3pt}r{3pt}){10-12} \cmidrule(l{3pt}r{3pt}){13-15} \cmidrule(l{3pt}r{3pt}){16-18} \cmidrule(l{3pt}r{3pt}){19-21}
 & Participants & \% Positive & M & CI & SD & M & CI & SD & M & CI & SD & M & CI & SD & M & CI & SD & M & CI & SD\\
\midrule
\textbf{\cellcolor{gray!6}{Mean Positive}} & \cellcolor{gray!6}{254 (51\%)} & \cellcolor{gray!6}{100.0} & \cellcolor{gray!6}{3.40} & \cellcolor{gray!6}{{}[3.3, 3.5]} & \cellcolor{gray!6}{1.63} & \cellcolor{gray!6}{4.07} & \cellcolor{gray!6}{{}[3.98, 4.15]} & \cellcolor{gray!6}{1.43} & \cellcolor{gray!6}{4.08} & \cellcolor{gray!6}{{}[4, 4.17]} & \cellcolor{gray!6}{1.39} & \cellcolor{gray!6}{4.45} & \cellcolor{gray!6}{{}[4.37, 4.53]} & \cellcolor{gray!6}{1.38} & \cellcolor{gray!6}{3.16} & \cellcolor{gray!6}{{}[3.07, 3.25]} & \cellcolor{gray!6}{1.52} & \cellcolor{gray!6}{3.98} & \cellcolor{gray!6}{{}[3.91, 4.05]} & \cellcolor{gray!6}{1.12}\\
\textbf{Mean Negative} & 243 (49\%) & 0.0 & 2.45 & {}[2.37, 2.54] & 1.46 & 3.12 & {}[3.03, 3.2] & 1.44 & 3.17 & {}[3.09, 3.25] & 1.34 & 3.30 & {}[3.21, 3.39] & 1.50 & 2.61 & {}[2.53, 2.7] & 1.46 & 3.03 & {}[2.96, 3.1] & 1.15\\
\midrule
\textbf{\cellcolor{gray!6}{High Quality}} & \cellcolor{gray!6}{165 (33.2\%)} & \cellcolor{gray!6}{74.5} & \cellcolor{gray!6}{4.37} & \cellcolor{gray!6}{{}[4.27, 4.46]} & \cellcolor{gray!6}{1.31} & \cellcolor{gray!6}{4.63} & \cellcolor{gray!6}{{}[4.54, 4.73]} & \cellcolor{gray!6}{1.33} & \cellcolor{gray!6}{4.64} & \cellcolor{gray!6}{{}[4.56, 4.72]} & \cellcolor{gray!6}{1.13} & \cellcolor{gray!6}{5.17} & \cellcolor{gray!6}{{}[5.09, 5.25]} & \cellcolor{gray!6}{1.07} & \cellcolor{gray!6}{3.97} & \cellcolor{gray!6}{{}[3.86, 4.08]} & \cellcolor{gray!6}{1.43} & \cellcolor{gray!6}{4.68} & \cellcolor{gray!6}{{}[4.62, 4.74]} & \cellcolor{gray!6}{0.82}\\
\textbf{Medium} & 146 (29.4\%) & 76.0 & 2.40 & {}[2.31, 2.5] & 1.23 & 3.56 & {}[3.47, 3.65] & 1.14 & 3.35 & {}[3.25, 3.45] & 1.30 & 3.60 & {}[3.51, 3.7] & 1.23 & 2.31 & {}[2.22, 2.39] & 1.04 & 3.22 & {}[3.15, 3.28] & 0.77\\
\textbf{\cellcolor{gray!6}{External}} & \cellcolor{gray!6}{58 (11.7\%)} & \cellcolor{gray!6}{19.0} & \cellcolor{gray!6}{2.40} & \cellcolor{gray!6}{{}[2.23, 2.56]} & \cellcolor{gray!6}{1.31} & \cellcolor{gray!6}{2.77} & \cellcolor{gray!6}{{}[2.61, 2.93]} & \cellcolor{gray!6}{1.28} & \cellcolor{gray!6}{3.77} & \cellcolor{gray!6}{{}[3.62, 3.93]} & \cellcolor{gray!6}{1.26} & \cellcolor{gray!6}{3.51} & \cellcolor{gray!6}{{}[3.31, 3.71]} & \cellcolor{gray!6}{1.58} & \cellcolor{gray!6}{3.24} & \cellcolor{gray!6}{{}[3.03, 3.44]} & \cellcolor{gray!6}{1.63} & \cellcolor{gray!6}{3.32} & \cellcolor{gray!6}{{}[3.18, 3.47]} & \cellcolor{gray!6}{1.14}\\
\textbf{Amotivated I.M.} & 71 (14.3\%) & 9.9 & 2.14 & {}[2.01, 2.28] & 1.20 & 3.48 & {}[3.34, 3.62] & 1.23 & 2.73 & {}[2.61, 2.86] & 1.11 & 2.96 & {}[2.85, 3.07] & 1.01 & 2.26 & {}[2.14, 2.38] & 1.06 & 2.68 & {}[2.6, 2.76] & 0.74\\
\textbf{\cellcolor{gray!6}{Amotivated}} & \cellcolor{gray!6}{57 (11.5\%)} & \cellcolor{gray!6}{3.5} & \cellcolor{gray!6}{1.65} & \cellcolor{gray!6}{{}[1.53, 1.78]} & \cellcolor{gray!6}{0.98} & \cellcolor{gray!6}{1.62} & \cellcolor{gray!6}{{}[1.57, 1.67]} & \cellcolor{gray!6}{0.39} & \cellcolor{gray!6}{2.40} & \cellcolor{gray!6}{{}[2.26, 2.54]} & \cellcolor{gray!6}{1.11} & \cellcolor{gray!6}{2.40} & \cellcolor{gray!6}{{}[2.24, 2.56]} & \cellcolor{gray!6}{1.26} & \cellcolor{gray!6}{1.70} & \cellcolor{gray!6}{{}[1.58, 1.82]} & \cellcolor{gray!6}{0.96} & \cellcolor{gray!6}{2.09} & \cellcolor{gray!6}{{}[1.99, 2.19]} & \cellcolor{gray!6}{0.79}\\
\bottomrule
\end{tabular}}
\end{table}
\end{landscape}

\begin{table}

\caption{\label{tab:wald_classes}Wald tests for comparison between classes}
\centering
\begin{tabular}[t]{lll}
\toprule
Profile 1 & Profile 2 & Wald statistic\\
\midrule
\cellcolor{gray!6}{High Quality} & \cellcolor{gray!6}{External} & \cellcolor{gray!6}{$\chi^2(6)=227.1, p < 0.0001$}\\
High Quality & Amotivated I.M. & $\chi^2(6)=507.2, p < 0.0001$\\
\cellcolor{gray!6}{High Quality} & \cellcolor{gray!6}{Amotivated} & \cellcolor{gray!6}{$\chi^2(6)=675, p < 0.0001$}\\
Medium & High Quality & $\chi^2(6)=320, p < 0.0001$\\
\cellcolor{gray!6}{Medium} & \cellcolor{gray!6}{External} & \cellcolor{gray!6}{$\chi^2(6)=111.1, p < 0.0001$}\\
Medium & Amotivated I.M. & $\chi^2(6)=216.3, p < 0.0001$\\
\cellcolor{gray!6}{Medium} & \cellcolor{gray!6}{Amotivated} & \cellcolor{gray!6}{$\chi^2(6)=430.5, p < 0.0001$}\\
Amotivated I.M. & External & $\chi^2(6)=91.1, p < 0.0001$\\
\cellcolor{gray!6}{Amotivated} & \cellcolor{gray!6}{External} & \cellcolor{gray!6}{$\chi^2(6)=92.5, p < 0.0001$}\\
Amotivated & Amotivated I.M. & $\chi^2(6)=178.7, p < 0.0001$\\
\bottomrule
\end{tabular}
\end{table}

\begin{table}

\caption{\label{tab:hypothesis_results}Wald tests for comparing variables between profiles}
\centering
\resizebox{0.73\linewidth}{!}{
\begin{tabular}[t]{lllllll}
\toprule
Variable & Profile 1 & Profile 2 & $\chi^2$ & p value & p value (adjusted) & significance\\
\midrule
\cellcolor{gray!6}{Autonomy} & \cellcolor{gray!6}{High Quality} & \cellcolor{gray!6}{External} & \cellcolor{gray!6}{102.9} & \cellcolor{gray!6}{0.00000} & \cellcolor{gray!6}{0.00000} & \cellcolor{gray!6}{***}\\
Autonomy & High Quality & Amotivated I.M. & 224.7 & 0.00000 & 0.00000 & ***\\
\cellcolor{gray!6}{Autonomy} & \cellcolor{gray!6}{High Quality} & \cellcolor{gray!6}{Amotivated} & \cellcolor{gray!6}{473.6} & \cellcolor{gray!6}{0.00000} & \cellcolor{gray!6}{0.00000} & \cellcolor{gray!6}{***}\\
Autonomy & Medium & High Quality & 153.5 & 0.00000 & 0.00000 & ***\\
\cellcolor{gray!6}{Autonomy} & \cellcolor{gray!6}{Medium} & \cellcolor{gray!6}{External} & \cellcolor{gray!6}{10.0} & \cellcolor{gray!6}{0.00158} & \cellcolor{gray!6}{0.02217} & \cellcolor{gray!6}{*}\\
Autonomy & Medium & Amotivated I.M. & 27.0 & 0.00000 & 0.00001 & ***\\
\cellcolor{gray!6}{Autonomy} & \cellcolor{gray!6}{Medium} & \cellcolor{gray!6}{Amotivated} & \cellcolor{gray!6}{103.6} & \cellcolor{gray!6}{0.00000} & \cellcolor{gray!6}{0.00000} & \cellcolor{gray!6}{***}\\
Autonomy & Amotivated I.M. & External & 0.6 & 0.45798 & 1.00000 & ns\\
\cellcolor{gray!6}{Autonomy} & \cellcolor{gray!6}{Amotivated} & \cellcolor{gray!6}{External} & \cellcolor{gray!6}{14.1} & \cellcolor{gray!6}{0.00017} & \cellcolor{gray!6}{0.00428} & \cellcolor{gray!6}{**}\\
Autonomy & Amotivated & Amotivated I.M. & 14.1 & 0.00017 & 0.00428 & **\\
\cellcolor{gray!6}{Competence} & \cellcolor{gray!6}{High Quality} & \cellcolor{gray!6}{External} & \cellcolor{gray!6}{0.9} & \cellcolor{gray!6}{0.33209} & \cellcolor{gray!6}{1.00000} & \cellcolor{gray!6}{ns}\\
Competence & High Quality & Amotivated I.M. & 40.1 & 0.00000 & 0.00000 & ***\\
\cellcolor{gray!6}{Competence} & \cellcolor{gray!6}{High Quality} & \cellcolor{gray!6}{Amotivated} & \cellcolor{gray!6}{122.1} & \cellcolor{gray!6}{0.00000} & \cellcolor{gray!6}{0.00000} & \cellcolor{gray!6}{***}\\
Competence & Medium & High Quality & 32.1 & 0.00000 & 0.00000 & ***\\
\cellcolor{gray!6}{Competence} & \cellcolor{gray!6}{Medium} & \cellcolor{gray!6}{External} & \cellcolor{gray!6}{7.7} & \cellcolor{gray!6}{0.00564} & \cellcolor{gray!6}{0.06774} & \cellcolor{gray!6}{ns}\\
Competence & Medium & Amotivated I.M. & 1.3 & 0.26218 & 1.00000 & ns\\
\cellcolor{gray!6}{Competence} & \cellcolor{gray!6}{Medium} & \cellcolor{gray!6}{Amotivated} & \cellcolor{gray!6}{29.2} & \cellcolor{gray!6}{0.00000} & \cellcolor{gray!6}{0.00000} & \cellcolor{gray!6}{***}\\
Competence & Amotivated I.M. & External & 11.9 & 0.00056 & 0.01110 & *\\
\cellcolor{gray!6}{Competence} & \cellcolor{gray!6}{Amotivated} & \cellcolor{gray!6}{External} & \cellcolor{gray!6}{41.6} & \cellcolor{gray!6}{0.00000} & \cellcolor{gray!6}{0.00000} & \cellcolor{gray!6}{***}\\
Competence & Amotivated & Amotivated I.M. & 15.2 & 0.00010 & 0.00255 & **\\
\cellcolor{gray!6}{Relatedness} & \cellcolor{gray!6}{High Quality} & \cellcolor{gray!6}{External} & \cellcolor{gray!6}{28.9} & \cellcolor{gray!6}{0.00000} & \cellcolor{gray!6}{0.00000} & \cellcolor{gray!6}{***}\\
Relatedness & High Quality & Amotivated I.M. & 381.1 & 0.00000 & 0.00000 & ***\\
\cellcolor{gray!6}{Relatedness} & \cellcolor{gray!6}{High Quality} & \cellcolor{gray!6}{Amotivated} & \cellcolor{gray!6}{116.3} & \cellcolor{gray!6}{0.00000} & \cellcolor{gray!6}{0.00000} & \cellcolor{gray!6}{***}\\
Relatedness & Medium & High Quality & 78.8 & 0.00000 & 0.00000 & ***\\
\cellcolor{gray!6}{Relatedness} & \cellcolor{gray!6}{Medium} & \cellcolor{gray!6}{External} & \cellcolor{gray!6}{1.7} & \cellcolor{gray!6}{0.18833} & \cellcolor{gray!6}{1.00000} & \cellcolor{gray!6}{ns}\\
Relatedness & Medium & Amotivated I.M. & 69.6 & 0.00000 & 0.00000 & ***\\
\cellcolor{gray!6}{Relatedness} & \cellcolor{gray!6}{Medium} & \cellcolor{gray!6}{Amotivated} & \cellcolor{gray!6}{10.8} & \cellcolor{gray!6}{0.00099} & \cellcolor{gray!6}{0.01629} & \cellcolor{gray!6}{*}\\
Relatedness & Amotivated I.M. & External & 61.2 & 0.00000 & 0.00000 & ***\\
\cellcolor{gray!6}{Relatedness} & \cellcolor{gray!6}{Amotivated} & \cellcolor{gray!6}{External} & \cellcolor{gray!6}{15.6} & \cellcolor{gray!6}{0.00008} & \cellcolor{gray!6}{0.00211} & \cellcolor{gray!6}{**}\\
Relatedness & Amotivated & Amotivated I.M. & 11.1 & 0.00085 & 0.01528 & *\\
\cellcolor{gray!6}{PANAS} & \cellcolor{gray!6}{High Quality} & \cellcolor{gray!6}{External} & \cellcolor{gray!6}{167.8} & \cellcolor{gray!6}{0.00000} & \cellcolor{gray!6}{0.00000} & \cellcolor{gray!6}{***}\\
PANAS & High Quality & Amotivated I.M. & 538.9 & 0.00000 & 0.00000 & ***\\
\cellcolor{gray!6}{PANAS} & \cellcolor{gray!6}{High Quality} & \cellcolor{gray!6}{Amotivated} & \cellcolor{gray!6}{1246.8} & \cellcolor{gray!6}{0.00000} & \cellcolor{gray!6}{0.00000} & \cellcolor{gray!6}{***}\\
PANAS & Medium & High Quality & 104.4 & 0.00000 & 0.00000 & ***\\
\cellcolor{gray!6}{PANAS} & \cellcolor{gray!6}{Medium} & \cellcolor{gray!6}{External} & \cellcolor{gray!6}{61.4} & \cellcolor{gray!6}{0.00000} & \cellcolor{gray!6}{0.00000} & \cellcolor{gray!6}{***}\\
PANAS & Medium & Amotivated I.M. & 219.3 & 0.00000 & 0.00000 & ***\\
\cellcolor{gray!6}{PANAS} & \cellcolor{gray!6}{Medium} & \cellcolor{gray!6}{Amotivated} & \cellcolor{gray!6}{608.8} & \cellcolor{gray!6}{0.00000} & \cellcolor{gray!6}{0.00000} & \cellcolor{gray!6}{***}\\
PANAS & Amotivated I.M. & External & 0.8 & 0.37454 & 1.00000 & ns\\
\cellcolor{gray!6}{PANAS} & \cellcolor{gray!6}{Amotivated} & \cellcolor{gray!6}{External} & \cellcolor{gray!6}{10.9} & \cellcolor{gray!6}{0.00096} & \cellcolor{gray!6}{0.01629} & \cellcolor{gray!6}{*}\\
PANAS & Amotivated & Amotivated I.M. & 17.3 & 0.00003 & 0.00091 & ***\\
\cellcolor{gray!6}{Self-Attribution} & \cellcolor{gray!6}{High Quality} & \cellcolor{gray!6}{External} & \cellcolor{gray!6}{7.5} & \cellcolor{gray!6}{0.00623} & \cellcolor{gray!6}{0.06857} & \cellcolor{gray!6}{ns}\\
Self-Attribution & High Quality & Amotivated I.M. & 2.7 & 0.09968 & 0.99679 & ns\\
\cellcolor{gray!6}{Self-Attribution} & \cellcolor{gray!6}{High Quality} & \cellcolor{gray!6}{Amotivated} & \cellcolor{gray!6}{41.8} & \cellcolor{gray!6}{0.00000} & \cellcolor{gray!6}{0.00000} & \cellcolor{gray!6}{***}\\
Self-Attribution & Medium & High Quality & 11.7 & 0.00061 & 0.01159 & *\\
\cellcolor{gray!6}{Self-Attribution} & \cellcolor{gray!6}{Medium} & \cellcolor{gray!6}{External} & \cellcolor{gray!6}{0.1} & \cellcolor{gray!6}{0.82238} & \cellcolor{gray!6}{1.00000} & \cellcolor{gray!6}{ns}\\
Self-Attribution & Medium & Amotivated I.M. & 1.5 & 0.21364 & 1.00000 & ns\\
\cellcolor{gray!6}{Self-Attribution} & \cellcolor{gray!6}{Medium} & \cellcolor{gray!6}{Amotivated} & \cellcolor{gray!6}{13.9} & \cellcolor{gray!6}{0.00020} & \cellcolor{gray!6}{0.00453} & \cellcolor{gray!6}{**}\\
Self-Attribution & Amotivated I.M. & External & 1.5 & 0.22218 & 1.00000 & ns\\
\cellcolor{gray!6}{Self-Attribution} & \cellcolor{gray!6}{Amotivated} & \cellcolor{gray!6}{External} & \cellcolor{gray!6}{9.2} & \cellcolor{gray!6}{0.00242} & \cellcolor{gray!6}{0.03143} & \cellcolor{gray!6}{*}\\
Self-Attribution & Amotivated & Amotivated I.M. & 20.7 & 0.00001 & 0.00016 & ***\\
\cellcolor{gray!6}{UMUX} & \cellcolor{gray!6}{High Quality} & \cellcolor{gray!6}{External} & \cellcolor{gray!6}{74.9} & \cellcolor{gray!6}{0.00000} & \cellcolor{gray!6}{0.00000} & \cellcolor{gray!6}{***}\\
UMUX & High Quality & Amotivated I.M. & 12.3 & 0.00044 & 0.00932 & **\\
\cellcolor{gray!6}{UMUX} & \cellcolor{gray!6}{High Quality} & \cellcolor{gray!6}{Amotivated} & \cellcolor{gray!6}{360.7} & \cellcolor{gray!6}{0.00000} & \cellcolor{gray!6}{0.00000} & \cellcolor{gray!6}{***}\\
UMUX & Medium & High Quality & 0.1 & 0.69862 & 1.00000 & ns\\
\cellcolor{gray!6}{UMUX} & \cellcolor{gray!6}{Medium} & \cellcolor{gray!6}{External} & \cellcolor{gray!6}{71.9} & \cellcolor{gray!6}{0.00000} & \cellcolor{gray!6}{0.00000} & \cellcolor{gray!6}{***}\\
UMUX & Medium & Amotivated I.M. & 10.3 & 0.00132 & 0.01987 & *\\
\cellcolor{gray!6}{UMUX} & \cellcolor{gray!6}{Medium} & \cellcolor{gray!6}{Amotivated} & \cellcolor{gray!6}{353.8} & \cellcolor{gray!6}{0.00000} & \cellcolor{gray!6}{0.00000} & \cellcolor{gray!6}{***}\\
UMUX & Amotivated I.M. & External & 38.6 & 0.00000 & 0.00000 & ***\\
\cellcolor{gray!6}{UMUX} & \cellcolor{gray!6}{Amotivated} & \cellcolor{gray!6}{External} & \cellcolor{gray!6}{12.9} & \cellcolor{gray!6}{0.00033} & \cellcolor{gray!6}{0.00732} & \cellcolor{gray!6}{**}\\
UMUX & Amotivated & Amotivated I.M. & 201.0 & 0.00000 & 0.00000 & ***\\
\bottomrule
\end{tabular}}
\end{table}

\begin{table}[]
\caption{\label{tab:themes}Frequencies for most prominent themes. Full coding including all coded extracts is provided in additional materials at the paper's \href{https://osf.io/eny6v/}{OSF repository}.}
\resizebox{0.95\linewidth}{!}{
\begin{tabular}{lllllllll}
Theme & Sub-theme & \cellcolor[HTML]{F2DCDB}\textbf{HQ} & \cellcolor[HTML]{F2DCDB}\textbf{Med} & \cellcolor[HTML]{F2DCDB}\textbf{Extr} & \cellcolor[HTML]{F2DCDB}\textbf{Amo IM} & \cellcolor[HTML]{F2DCDB}\textbf{Amo} & Positive & Negative \\
\cellcolor[HTML]{E6B8B7}Qualities of   the Tech          & \cellcolor[HTML]{E6B8B7}Negative evaluations                        & \cellcolor[HTML]{E6B8B7}21\%        & \cellcolor[HTML]{E6B8B7}19\%         & \cellcolor[HTML]{E6B8B7}66\%          & \cellcolor[HTML]{E6B8B7}28\%            & \cellcolor[HTML]{E6B8B7}75\%         & \cellcolor[HTML]{E6B8B7}65\%     & \cellcolor[HTML]{E6B8B7}2\%      \\
\cellcolor[HTML]{E6B8B7}Qualities of   the Tech          & \cellcolor[HTML]{F2DCDB}Positive evaluations                        & \cellcolor[HTML]{F2DCDB}64\%        & \cellcolor[HTML]{F2DCDB}62\%         & \cellcolor[HTML]{F2DCDB}22\%          & \cellcolor[HTML]{F2DCDB}8\%             & \cellcolor[HTML]{F2DCDB}30\%         & \cellcolor[HTML]{F2DCDB}23\%     & \cellcolor[HTML]{F2DCDB}69\%     \\
\cellcolor[HTML]{E6B8B7}Qualities of   the Tech          & \cellcolor[HTML]{F2DCDB}Positive evaluations   (exceptional)        & \cellcolor[HTML]{F2DCDB}19\%        & \cellcolor[HTML]{F2DCDB}16\%         & \cellcolor[HTML]{F2DCDB}5\%           & \cellcolor[HTML]{F2DCDB}0\%             & \cellcolor[HTML]{F2DCDB}2\%          & \cellcolor[HTML]{F2DCDB}3\%      & \cellcolor[HTML]{F2DCDB}20\%     \\
\cellcolor[HTML]{D8E4BC}Good or Bad   For People         & \cellcolor[HTML]{D8E4BC}Good for people                             & \cellcolor[HTML]{D8E4BC}27\%        & \cellcolor[HTML]{D8E4BC}18\%         & \cellcolor[HTML]{D8E4BC}5\%           & \cellcolor[HTML]{D8E4BC}18\%            & \cellcolor[HTML]{D8E4BC}7\%          & \cellcolor[HTML]{D8E4BC}9\%      & \cellcolor[HTML]{D8E4BC}27\%     \\
\cellcolor[HTML]{D8E4BC}Good or Bad   For People         & \cellcolor[HTML]{EBF1DE}Bad for people                              & \cellcolor[HTML]{EBF1DE}8\%         & \cellcolor[HTML]{EBF1DE}12\%         & \cellcolor[HTML]{EBF1DE}24\%          & \cellcolor[HTML]{EBF1DE}61\%            & \cellcolor[HTML]{EBF1DE}35\%         & \cellcolor[HTML]{EBF1DE}42\%     & \cellcolor[HTML]{EBF1DE}2\%      \\

\cellcolor[HTML]{95B3D7}Purpose                          & \cellcolor[HTML]{95B3D7}Positive purpose   (Work and education)     & \cellcolor[HTML]{95B3D7}15\%        & \cellcolor[HTML]{95B3D7}18\%         & \cellcolor[HTML]{95B3D7}45\%          & \cellcolor[HTML]{95B3D7}3\%             & \cellcolor[HTML]{95B3D7}19\%         & \cellcolor[HTML]{95B3D7}19\%     & \cellcolor[HTML]{95B3D7}18\%     \\
\cellcolor[HTML]{95B3D7}Purpose                          & \cellcolor[HTML]{95B3D7}Positive purpose   (Entertainment)          & \cellcolor[HTML]{95B3D7}28\%        & \cellcolor[HTML]{95B3D7}34\%         & \cellcolor[HTML]{95B3D7}5\%           & \cellcolor[HTML]{95B3D7}49\%            & \cellcolor[HTML]{95B3D7}21\%         & \cellcolor[HTML]{95B3D7}23\%     & \cellcolor[HTML]{95B3D7}35\%     \\
\cellcolor[HTML]{95B3D7}Purpose                          & \cellcolor[HTML]{95B3D7}Positive purpose   (Money)                  & \cellcolor[HTML]{95B3D7}16\%        & \cellcolor[HTML]{95B3D7}17\%         & \cellcolor[HTML]{95B3D7}5\%           & \cellcolor[HTML]{95B3D7}6\%             & \cellcolor[HTML]{95B3D7}21\%         & \cellcolor[HTML]{95B3D7}11\%     & \cellcolor[HTML]{95B3D7}17\%     \\
\cellcolor[HTML]{95B3D7}Purpose                          & \cellcolor[HTML]{95B3D7}Positive purpose   (Health and Growth)      & \cellcolor[HTML]{95B3D7}42\%        & \cellcolor[HTML]{95B3D7}40\%         & \cellcolor[HTML]{95B3D7}16\%          & \cellcolor[HTML]{95B3D7}20\%            & \cellcolor[HTML]{95B3D7}16\%         & \cellcolor[HTML]{95B3D7}18\%     & \cellcolor[HTML]{95B3D7}46\%     \\
\cellcolor[HTML]{95B3D7}Purpose                          & \cellcolor[HTML]{95B3D7}Positive purpose   (Passions and Interests) & \cellcolor[HTML]{95B3D7}30\%        & \cellcolor[HTML]{95B3D7}25\%         & \cellcolor[HTML]{95B3D7}7\%           & \cellcolor[HTML]{95B3D7}7\%             & \cellcolor[HTML]{95B3D7}4\%          & \cellcolor[HTML]{95B3D7}9\%      & \cellcolor[HTML]{95B3D7}30\%     \\
\cellcolor[HTML]{95B3D7}Purpose                          & \cellcolor[HTML]{95B3D7}Positive purpose   (Connecting)             & \cellcolor[HTML]{95B3D7}32\%        & \cellcolor[HTML]{95B3D7}33\%         & \cellcolor[HTML]{95B3D7}40\%          & \cellcolor[HTML]{95B3D7}28\%            & \cellcolor[HTML]{95B3D7}14\%         & \cellcolor[HTML]{95B3D7}24\%     & \cellcolor[HTML]{95B3D7}37\%     \\
\cellcolor[HTML]{95B3D7}Purpose                          & \cellcolor[HTML]{B8CCE4}Attitudes   (Self-driving)                  & \cellcolor[HTML]{B8CCE4}48\%        & \cellcolor[HTML]{B8CCE4}38\%         & \cellcolor[HTML]{B8CCE4}24\%          & \cellcolor[HTML]{B8CCE4}8\%             & \cellcolor[HTML]{B8CCE4}26\%         & \cellcolor[HTML]{B8CCE4}21\%     & \cellcolor[HTML]{B8CCE4}47\%     \\
\cellcolor[HTML]{95B3D7}Purpose                          & \cellcolor[HTML]{B8CCE4}Attitudes (Compulsion   and bad babits)     & \cellcolor[HTML]{B8CCE4}3\%         & \cellcolor[HTML]{B8CCE4}8\%          & \cellcolor[HTML]{B8CCE4}10\%          & \cellcolor[HTML]{B8CCE4}63\%            & \cellcolor[HTML]{B8CCE4}18\%         & \cellcolor[HTML]{B8CCE4}29\%     & \cellcolor[HTML]{B8CCE4}2\%      \\
\cellcolor[HTML]{95B3D7}Purpose                          & \cellcolor[HTML]{B8CCE4}Attitudes (Escapism)                        & \cellcolor[HTML]{B8CCE4}8\%         & \cellcolor[HTML]{B8CCE4}14\%         & \cellcolor[HTML]{B8CCE4}2\%           & \cellcolor[HTML]{B8CCE4}32\%            & \cellcolor[HTML]{B8CCE4}5\%          & \cellcolor[HTML]{B8CCE4}12\%     & \cellcolor[HTML]{B8CCE4}13\%     \\
\cellcolor[HTML]{95B3D7}Purpose                          & \cellcolor[HTML]{B8CCE4}Attitudes (Not my   choice)                 & \cellcolor[HTML]{B8CCE4}5\%         & \cellcolor[HTML]{B8CCE4}10\%         & \cellcolor[HTML]{B8CCE4}55\%          & \cellcolor[HTML]{B8CCE4}3\%             & \cellcolor[HTML]{B8CCE4}30\%         & \cellcolor[HTML]{B8CCE4}24\%     & \cellcolor[HTML]{B8CCE4}6\%      \\
\cellcolor[HTML]{95B3D7}Purpose                          & \cellcolor[HTML]{B8CCE4}Attitudes (Intend to   quit)                & \cellcolor[HTML]{B8CCE4}1\%         & \cellcolor[HTML]{B8CCE4}1\%          & \cellcolor[HTML]{B8CCE4}2\%           & \cellcolor[HTML]{B8CCE4}8\%             & \cellcolor[HTML]{B8CCE4}11\%         & \cellcolor[HTML]{B8CCE4}7\%      & \cellcolor[HTML]{B8CCE4}0\%      \\

\cellcolor[HTML]{FABF8F}Values, identity, and Emotions & \cellcolor[HTML]{FABF8F}Positive connection   to Identity           & \cellcolor[HTML]{FABF8F}19\%        & \cellcolor[HTML]{FABF8F}12\%         & \cellcolor[HTML]{FABF8F}3\%           & \cellcolor[HTML]{FABF8F}7\%             & \cellcolor[HTML]{FABF8F}4\%          & \cellcolor[HTML]{FABF8F}6\%      & \cellcolor[HTML]{FABF8F}17\%     \\
\cellcolor[HTML]{FABF8F}Values, identity, and Emotions & \cellcolor[HTML]{FCD5B4}Positive emotional   associations           & \cellcolor[HTML]{FCD5B4}14\%        & \cellcolor[HTML]{FCD5B4}11\%         & \cellcolor[HTML]{FCD5B4}0\%           & \cellcolor[HTML]{FCD5B4}1\%             & \cellcolor[HTML]{FCD5B4}2\%          & \cellcolor[HTML]{FCD5B4}3\%      & \cellcolor[HTML]{FCD5B4}13\%     \\
\cellcolor[HTML]{FABF8F}Values, identity, and Emotions & \cellcolor[HTML]{FABF8F}Value disagreement                          & \cellcolor[HTML]{FABF8F}4\%         & \cellcolor[HTML]{FABF8F}10\%         & \cellcolor[HTML]{FABF8F}16\%          & \cellcolor[HTML]{FABF8F}38\%            & \cellcolor[HTML]{FABF8F}23\%         & \cellcolor[HTML]{FABF8F}26\%     & \cellcolor[HTML]{FABF8F}2\%      \\
\cellcolor[HTML]{FABF8F}Values, identity, and Emotions & \cellcolor[HTML]{FABF8F}Value disagreement   (Maker is bad)       & \cellcolor[HTML]{FABF8F}1\%         & \cellcolor[HTML]{FABF8F}4\%          & \cellcolor[HTML]{FABF8F}3\%           & \cellcolor[HTML]{FABF8F}4\%             & \cellcolor[HTML]{FABF8F}5\%          & \cellcolor[HTML]{FABF8F}6\%      & \cellcolor[HTML]{FABF8F}0\%      \\
\cellcolor[HTML]{FABF8F}Values, identity, and Emotions & \cellcolor[HTML]{FABF8F}Value disagreement   (Users are bad)        & \cellcolor[HTML]{FABF8F}1\%         & \cellcolor[HTML]{FABF8F}0\%          & \cellcolor[HTML]{FABF8F}2\%           & \cellcolor[HTML]{FABF8F}11\%            & \cellcolor[HTML]{FABF8F}5\%          & \cellcolor[HTML]{FABF8F}5\%      & \cellcolor[HTML]{FABF8F}0\%      \\
\cellcolor[HTML]{FABF8F}Values, identity, and Emotions & \cellcolor[HTML]{FCD5B4}Value agreement                             & \cellcolor[HTML]{FCD5B4}6\%         & \cellcolor[HTML]{FCD5B4}7\%          & \cellcolor[HTML]{FCD5B4}2\%           & \cellcolor[HTML]{FCD5B4}3\%             & \cellcolor[HTML]{FCD5B4}0\%          & \cellcolor[HTML]{FCD5B4}2\%      & \cellcolor[HTML]{FCD5B4}7\%      \\
\cellcolor[HTML]{FABF8F}Values, identity, and Emotions & \cellcolor[HTML]{FCD5B4}Value agreement   (Ethics)                  & \cellcolor[HTML]{FCD5B4}2\%         & \cellcolor[HTML]{FCD5B4}1\%          & \cellcolor[HTML]{FCD5B4}0\%           & \cellcolor[HTML]{FCD5B4}1\%             & \cellcolor[HTML]{FCD5B4}0\%          & \cellcolor[HTML]{FCD5B4}0\%      & \cellcolor[HTML]{FCD5B4}2\%      \\
\rowcolor[HTML]{595959} 
\cellcolor[HTML]{FABF8F}Values, identity, and Emotions & \cellcolor[HTML]{FCD5B4}Value agreement   (Maker is good)         & \cellcolor[HTML]{FCD5B4}2\%         & \cellcolor[HTML]{FCD5B4}0\%                                  & \cellcolor[HTML]{FCD5B4}0\%                                   & \cellcolor[HTML]{FCD5B4}0\%                                     & \cellcolor[HTML]{FCD5B4}0\%                                  & \cellcolor[HTML]{FCD5B4}0\%                              & \cellcolor[HTML]{FCD5B4}1\%      \\
\cellcolor[HTML]{FABF8F}Values, identity, and Emotions & \cellcolor[HTML]{FABF8F}No deeper meaning                           & \cellcolor[HTML]{FABF8F}4\%         & \cellcolor[HTML]{FABF8F}18\%         & \cellcolor[HTML]{FABF8F}14\%          & \cellcolor[HTML]{FABF8F}31\%            & \cellcolor[HTML]{FABF8F}33\%         & \cellcolor[HTML]{FABF8F}21\%     & \cellcolor[HTML]{FABF8F}12\%     \\
\cellcolor[HTML]{FABF8F}Values, identity, and Emotions & \cellcolor[HTML]{FABF8F}No deeper meaning   (actually meaningless)  & \cellcolor[HTML]{FABF8F}0\%         & \cellcolor[HTML]{FABF8F}0\%          & \cellcolor[HTML]{FABF8F}5\%           & \cellcolor[HTML]{FABF8F}13\%            & \cellcolor[HTML]{FABF8F}9\%          & \cellcolor[HTML]{FABF8F}7\%      & \cellcolor[HTML]{FABF8F}0\%     
\end{tabular}
}
\end{table}

\end{document}